\documentclass[hidelinks,journal]{IEEEtran}
\IEEEoverridecommandlockouts

\usepackage{cite}
\usepackage{amsmath,amssymb,amsfonts}
\usepackage{graphicx}
\usepackage{textcomp}
\usepackage[dvipsnames]{xcolor}
\usepackage{orcidlink}
\usepackage[acronym,shortcuts]{glossaries}
\usepackage{bm}
\usepackage{caption}
\usepackage{subcaption}
\usepackage{relsize}
\usepackage{soul}
\usepackage{algorithm, algpseudocode}
\usepackage{outlines}
\usepackage{lipsum}
\usepackage{scalefnt}
\usepackage{dblfloatfix}
\usepackage[bottom]{footmisc}
\usepackage{comment}

\newcommand{\trans}[0]{^{\mathsf{T}}}
\newcommand{\transs}[0]{^{\!\mathsf{T}}}

\newacronym{WSN}{WSN}{wireless sensor network}
\newacronym{WLS}{WLS}{weighted least squares}
\newacronym{DAC}{DAC}{divide and conquer}
\newacronym{TDOA}{TDOA}{time difference of arrival}
\newacronym{TOA}{TOA}{time of arrival}
\newacronym{IMU}{IMU}{inertial measurement unit}
\newacronym{MFB}{MFB}{matched filter bound}
\newacronym{BP}{BP}{belief propagation}
\newacronym{RMSE}{RMSE}{root-mean-squared-error}
\newacronym{VR}{VR}{virtual reality}
\newacronym{XR}{XR}{extended reality}
\newacronym{SotA}{SotA}{state-of-the-art}
\newacronym{MSE}{MSE}{mean-squared-error}
\newacronym{PDF}{PDF}{probability density function}
\newacronym{SGA}{SGA}{scalar Gaussian approximation}
\newacronym{ML}{ML}{machine learning}
\newacronym{IC}{IC}{interference cancellation}
\newacronym{GaBP}{GaBP}{Gaussian belief propagation}
\newacronym{RBL}{RBL}{rigid body localization}
\newacronym{AWGN}{AWGN}{additive white Gaussian noise}
\newacronym{3D}{3D}{three-dimensional}
\newacronym{6D}{6D}{sixth-dimensional}
\newacronym{IoT}{IoT}{Internet-of-Things}
\newacronym{i.i.d.}{i.i.d.}{independent and identically distributed}
\newacronym{SVD}{SVD}{singular value decomposition}
\newacronym{wlg}{w.l.g.}{without loss of generality}
\newacronym{sIC}{soft-IC}{soft interference cancellation}
\newacronym{FDOA}{FDOA}{frequency difference of arrival}
\newacronym{SDR}{SDR}{semidefinite relaxation}

\newacronym{MDS}{MDS}{multidimensional scaling}

\newacronym{ToA}{ToA}{time of arrival}
\newacronym{RSSI}{RSSI}{received signal strength indicator}
\newacronym{MIMO}{MIMO}{multiple-input multiple-output}

\title{ $~$ \\[-3ex] {\normalsize{\textbf{\color{red}Please find the official IEEE \underline{published} version of this article on IEEE Xplore \href{https://ieeexplore.ieee.org/document/11181201}{{\color{blue}[here]}}} \textbf{\color{red}and cite as:} \\} 

\color{cyan} N F\"uhrling \emph{et al.}, ``6D Rigid Body Localization and Velocity Estimation via Gaussian Belief Propagation,"\\[-4ex]

\emph{in IEEE Transactions on Signal Processing}, vol. 73, pp. 3902-3917, Sept. 2025}\\[-0.5ex]

6D Rigid Body Localization and Velocity Estimation via Gaussian Belief Propagation\vspace{-.5ex}}

\author{Niclas~F\"uhrling\textsuperscript{\orcidlink{0000-0003-1942-8691}},~\IEEEmembership{Graduate Student Member,~IEEE}, Volodymyr~Vizitiv\textsuperscript{\orcidlink{0009-0002-9451-8216}},~\IEEEmembership{Graduate Student Member,~IEEE},\\
Kuranage Roche Rayan Ranasinghe\textsuperscript{\orcidlink{0000-0002-6834-8877}},~\IEEEmembership{Graduate Student Member,~IEEE}, Hyeon Seok Rou\textsuperscript{\orcidlink{0000-0003-3483-7629}}, \IEEEmembership{Member,~IEEE,}\\
Giuseppe Thadeu Freitas de Abreu\textsuperscript{\orcidlink{0000-0002-5018-8174}},~\IEEEmembership{Senior Member,~IEEE,}\\ David~Gonz{\'a}lez~G.\textsuperscript{\orcidlink{0000-0003-2090-8481}},~\IEEEmembership{Senior Member,~IEEE,} and Osvaldo~Gonsa\textsuperscript{\orcidlink{0000-0001-5452-8159}}

\thanks{N.~F\"uhrling, K.~R.~R.~Ranasinghe, H.~S.~Rou and G.~T.~F.~de~Abreu are with the School of Computer Science and Engineering, Constructor University, Campus Ring 1, 28759, Bremen, Germany {(e-mails: [nfuehrling, kranasinghe, hrou, gabreu]@constructor.university).}}
\thanks{V.~Vizitiv is with the George~R.~Brown School of Engineering and Computing, Rice University, 6100 Main Street, Houston, Texas, USA (email: vv38@rice.edu).}
\thanks{D.~Gonz{\'a}lez~G. and O.~Gonsa are with Continental Automotive Technologies GmbH, Guerickestrasse 7, 60488, Frankfurt am Main, Germany (e-mails: david.gonzalez.g@ieee.org, osvaldo.gonsa@continental-corporation.com).}
\thanks{Parts of this article have been presented at the 2025 IEEE Wireless Communications and Networking Conference (WCNC) \cite{Vizitiv_2024} (Corresponding author: V.~Vizitiv)}
\vspace{-5ex}
}

\begin{document}

\markboth{}{F\"uhrling \MakeLowercase{\textit{et al.}}: 6D Rigid Body Localization and Velocity Estimation via Gaussian Belief Propagation}
\maketitle 
\begin{abstract}
\begin{outline}
We propose a novel message-passing solution to the \ac{6D} moving \ac{RBL} problem, in which the \ac{3D} translation vector and rotation angles, as well as their corresponding translational and angular velocities, are all estimated by only utilizing the relative range and Doppler measurements between the ``anchor'' sensors located at an \ac{3D} (rigid body) observer and the ``target'' sensors of another rigid body. 
The proposed method is based on a bilinear \ac{GaBP} framework, employed to estimate the absolute sensor positions and velocities using a range- and Doppler-based received signal model, which is then utilized in the reconstruction of the \ac{RBL} transformation model, linearized under a small-angle approximation.
The method further incorporates a second bivariate \ac{GaBP} designed to directly estimate the \ac{3D} rotation angles and translation vectors, including an \ac{IC} refinement stage to improve the angle estimation performance, followed by the estimation of the angular and the translational velocities.
The effectiveness of the proposed method is verified via simulations, which confirms its improved performance compared to equivalent \ac{SotA} techniques.
\end{outline}
\end{abstract}

\glsresetall


\glsresetall

\vspace{-2.5ex}
\section{Introduction}
\label{sec:introduction} 

Recent years have seen great advancements in wireless sensor technologies, which are capable of detecting environmental parameters such as temperature, pressure, luminosity, humidity and the strength of electric signals, that are used in applications ranging from monitoring and control, to smart factories, to \ac{IoT}, to positioning systems \cite{Jamshed_SJ22, Kandris_ASI20, Lee_Sensor18} and more.
In fact, many \ac{WSN} applications either require or can be improved under the knowledge of accurate location information, such that the sensor location problem has been studied extensively \cite{PatwariSPM2005,GustafssonSPM2005}. 

More recent sensing-related applications such as \ac{VR}, \ac{XR}, robotics and autonomous vehicles, however, require not only precise location information of individual sensors, but also the orientation of the various sensors associated to a given body or object \cite{Yang_JMD09, Whittaker_IEEE06}, giving rise to a variation of the positioning problem known as \ac{RBL} \cite{EggertMVA1997, Diebel2006RigidBodyAttitude, ChenTSP2015, ZhaMRBL2021, Yu_ITJ23, fuehrling2024}, where the relative position of sets of sensors is fixed according to a \textit{rigid} conformation, whose translation and rotation (\textit{i.e.,} orientation or attitude) must be estimated.

Several effective strategies {\color{black}exist} to estimate the location and attitude of objects, including computer vision-based techniques, which focus on feature extraction and posture estimation \cite{Chaoyi_ICASSP21, Xiang_arxiv17}, usually based on image/video signals; and \ac{IMU}-based  techniques, which leverage information from accelerometers, gyroscopes, and magnetometers \cite{AghiliTM2013, Zhao_JPCS18}.
A problem of computer-vision approaches is, however, that they typically require high volumes of data and rely on high-complexity methods, which limits their application. 
In turn, \ac{IMU}-based methods are often unreliable or inaccurate, requiring frequent sensor re-calibrations, not to mention the aid of external radio technology, which to some extent defeats the self-reliant idea behind the approach.

In contrast to the latter, the \ac{RBL} concept considered here \cite{EggertMVA1997, Diebel2006RigidBodyAttitude, ChenTSP2015, ZhaMRBL2021, Yu_ITJ23, fuehrling2024} aims to generalize the traditional localization paradigm, and exploits range measurements between sensors on the rigid body and a set of \textit{anchor} sensors at known positions \cite{AlcocerCDC2008, SandPWCS2014}, to estimate the location, orientation (and possibly the shape) of rigid bodies.
Earlier \ac{SotA} contributions to such \ac{RBL} approaches are, however, based on either algebraic methods leveraging \ac{MDS} \cite{Pizzo2016} or \ac{SDR} \cite{Jiang2018, Wang2020}, both of which have a cubic computational complexity at least.
And although least squares methods \cite{Chepuri_TSP14, Zhou2019, Wang_2020_LS} have been used to lower the complexity of the \ac{RBL} problem, these techniques have in common the fact that they only take into account stationary scenarios, in the sense that they seek to estimate only the translation and the orientation ($i.e.$, the pose) of the rigid body, respectively defined as the distance and the rotation matrix of the rigid body relative to a given reference.

In contrast, in \ac{RBL} schemes suitable to moving scenarios one must, in addition to pose, also estimate the corresponding angular and translational velocities.
Unfortunately, existing moving \ac{RBL} solutions are either of low complexity but low precision, as is the case of the alternating minimization-based method proposed in \cite{Yu2023}, or of high precision but also of high complexity, such as the \ac{SDR} method of \cite{Jiang2019}.

An example of a well-balanced solution -- $i.e.$, namely, with both high precision and low complexity -- for both stationary and moving \ac{RBL} is, however, the method  proposed in \cite{ChenTSP2015}, where preliminary position estimates for each sensor is first obtained from range measurements using a least-squares formulation based on the linearized model described in \cite{MaICASSP2011}, followed by the extraction of rotation and translation parameters via \ac{SVD}-based analysis and finally a conformation-enforcing refinement based on an Euler angles formulation, solved via a \ac{WLS} approach.
However, this stationary \ac{RBL} technique is further extended into a two stage solution for the moving \ac{RBL} scenario, in which the preliminary range-based position estimation stage is upgraded to include also velocity estimates for each sensor obtained from Doppler measurements, followed by the estimation of angular and translational velocities using the linearized model described in \cite{HoTSP2004}.

The capitalizing feature of the method in \cite{ChenTSP2015} to handle both the stationary and moving \ac{RBL} problems similarly is thanks to the linearized error model described in \cite{MaICASSP2011} and \cite{HoTSP2004}.
However, the least-squares formulation and two stage approach adopted can limit performance \cite{Vaskevicius_LS2023} compared to message-passing methods \cite{FengAMP2022}, which are well known to achieve low complexity while maintaining a good performance especially in problems with coupled variables, which can be solved not only via alternating methods \cite{ZhuAltSAMP2024}, but also jointly via bilinear techniques \cite{ParkerBiGaMPI_2014, ParkerBiGaMPII_2014, Rou_TWC2024,Rou_Asilomar2022}.

{\color{black}
Among existing localization approaches based on message passing techniques, it should be noted that, to the best of the authors' knowledge, the only method specifically designed for \ac{RBL} is the work introduced in \cite{Vizitiv_2024}.
Nevertheless, several other message passing-based localization techniques, though not specifically designed for \ac{RBL}, are worth presenting. 
For example, the method presented in \cite{Yuan_2016} employs \ac{GaBP} for simultaneous localization and synchronization based on \ac{ToA} measurements. 
Additionally, in \cite{Jin_2020}, message passing algorithms are exploited to solve a probabilistic inference problem in \ac{RSSI}-based cooperative localization scenarios involving an unknown path loss exponent. 
Moreover, the approach described in \cite{Yu_2021} leverages \ac{GaBP} for distributed \ac{MIMO} radar localization, resulting in low computational complexity and reduced communication cost.

Motivated by the aforementioned studies, and the identification of the lack of message passing-based \ac{RBL} techniques -- a series of novel algorithms are proposed in this article to address both stationary and moving \ac{RBL} problems.}
First, two methods for the location and velocity estimation of sensors in \ac{3D} space are produced.
In particular, the techniques, labeled Algorithms \ref{alg:pos_GaBP} and \ref{alg:vel_GaBP}, respectively, are designed for the estimation of sensor positions from range measurements, as well as sensor velocities from Doppler measurements.
Then, two bivariate linear \ac{GaBP} algorithms are offered for the stationary and moving \ac{RBL} problems.
Those contributions, dubbed Algorithms \ref{alg:RBL_GaBP} and \ref{alg:RBL_GaBP_mov}, respectively, are capable of estimating the \ac{3D} rotation angles and translation distances, and respectively the \ac{3D} angular and translational velocities, utilizing range/Doppler measurements and the estimated sensor position and velocities from Algorithms \ref{alg:pos_GaBP} and \ref{alg:RBL_GaBP} by leveraging bilinear \ac{GaBP} over a small-angle modification of the linearized system models of \cite{HoTSP2004} and \cite{MaICASSP2011}, with the rigid body conformation constraint directly incorporated.
The bivariate methods in Algorithms \ref{alg:RBL_GaBP_mov} and \ref{alg:vel_GaBP}, respectively, are shown to outperform the \ac{SotA} two-stage \ac{RBL} methods in terms of parameter estimation performance, while retaining low computational complexity.

The contributions of the article are summarized as follows:
\begin{enumerate}
\item A new linear \ac{GaBP} algorithm for the estimation of \ac{3D} sensor positions via range measurements;
\item A linear \ac{GaBP} algorithm for the estimation of \ac{3D} sensor velocities via Doppler measurements;
\item A bivariate \ac{GaBP} algorithm for the estimation of \ac{3D} rotation angles and translation distances of a rigid body via range measurements and estimated sensor positions;

\item A bivariate \ac{GaBP} algorithm for the estimation of \ac{3D} angular and translational velocities of a moving rigid body via Doppler measurements and estimated velocities.
\end{enumerate}

The remainder of the article is structured as follows: the system and measurement models, as well as the formulation of the fundamental \ac{RBL} estimation problem, are described in Section \ref{sec:system_model}.
The proposed reformulation leveraging a linearized small angle model and the \ac{RBL} conformation constraints for stationary \ac{RBL} is presented in Section \ref{sec:proposed_stationary}, where the derivation of the corresponding message passing rules for the proposed \ac{GaBP} estimator of the translation and rotation parameters are also elaborated.
Section \ref{sec:moving_RBL} then presents the equivalent of the latter for the moving \ac{RBL} problem, and finally Section \ref{sec:performance_analysis} compares the performance of the proposed methods against \ac{SotA} techniques, both via numerical simulations and a complexity and convergence analysis.

\vspace{-1ex}
\section{Rigid Body Localization System Model}
\label{sec:system_model}

\subsection{Rigid Body System Model}
\label{sec:RBL_model}

\begin{figure}[t]
\centering
\includegraphics[width=0.8\columnwidth]{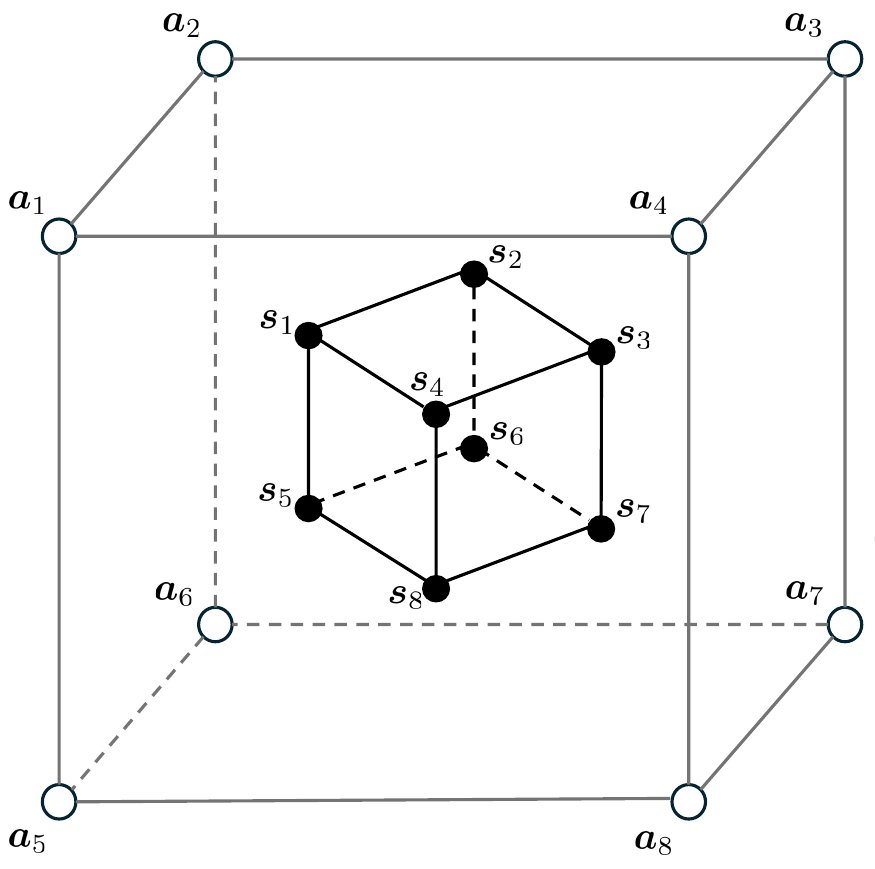}    
\vspace{-1ex}
\caption{An illustration of the anchor and rigid body sensor structures, in which a cubic rigid body ($N = 8$) is surrounded by a cubic deployment of anchors ($M = 8$).}
\label{fig:sys_mod_plot}
\end{figure}

Consider a scenario where a rigid body consisting of $N$ sensors is surrounded by a total of $M$ reference sensors (hereafter referred simply as anchors), as illustrated in Figure \ref{fig:sys_mod_plot}.
Each sensor and anchor is described by a $3 \times 1$ vector consisting of its  \textit{x-,y-,z}-coordinates in the \ac{3D} Euclidean space, respectively denoted by $\boldsymbol{c}_n \in \mathbb{R}^{3\times 1}$ for $n=\{1, \ldots, N\}$ and $\boldsymbol{a}_m \in \mathbb{R}^{3\times 1}$ for $m=\{1, \ldots, M\}$.
The initial sensor structure in the rigid body is consequently defined by a conformation matrix $\boldsymbol{C}=\left[\boldsymbol{c}_{1}, \boldsymbol{c}_{2}, \ldots, \boldsymbol{c}_{N}\right] \in \mathbb{R}^{3 \times N}$ at the reference frame (local axis) of the rigid body.

A transformation of the rigid body in \ac{3D} space can be fully defined by a translation and rotation, respectively described by the translation vector $\boldsymbol{t} \triangleq [t_x, t_y, t_z]\trans \in\mathbb{R}^{3\times 1}$ consisting of the translation distances in each axis, and a \ac{3D} rotation matrix\footnote{The rotation matrix $\boldsymbol{Q}$ is part of the special orthogonal group such that $SO(3)=\left\{\boldsymbol{Q} \in \mathbb{R}^{3 \times 3}: \boldsymbol{Q}\trans \boldsymbol{Q} = \mathbf{I}, ~\mathrm{det}(\boldsymbol{Q})=1\right\}$ \cite{Diebel2006RigidBodyAttitude} .} $\boldsymbol{Q} \in \mathbb{R}^{3\times 3}$ given by
\begin{align}
\label{eq:rotation_matrix}
\bm{Q} \triangleq\! \text{\scalebox{0.875}{$\overbrace{\left[
\begin{array}{@{}c@{\;\,}c@{\;\,}c@{}}
\cos\theta_z&-\sin\theta_z& 0\\
\sin\theta_z& \cos\theta_z& 0\\
0 	    & 0           & 1\\
\end{array}\right]}^{\triangleq \,\bm{Q}_z \in\, \mathbb{R}^{3\times 3}}\!\cdot\!
\overbrace{\left[
\begin{array}{@{}c@{\;\,}c@{\;\,}c@{}}
\cos\theta_y & 0           & \sin\theta_y\\
0			& 1			  & 0\\
-\sin\theta_y& 0 		  & \cos\theta_y\\
\end{array}\right]}^{\triangleq \,\bm{Q}_y \in\, \mathbb{R}^{3\times 3}}\!\cdot\!
\overbrace{\left[
\begin{array}{@{\,}c@{\;\,}c@{\;\,}c@{\!}}
1 			& 0			  & 0\\
0			& \cos\theta_x& -\sin\theta_x\\
0			& \sin\theta_x& \cos\theta_x\\
\end{array}\right]}^{\triangleq \bm{Q}_x \in\, \mathbb{R}^{3\times 3}}$}}\!, \\[-5ex] \nonumber 
\end{align}

\begin{figure}[t]
\centering
\includegraphics[width=1\columnwidth]{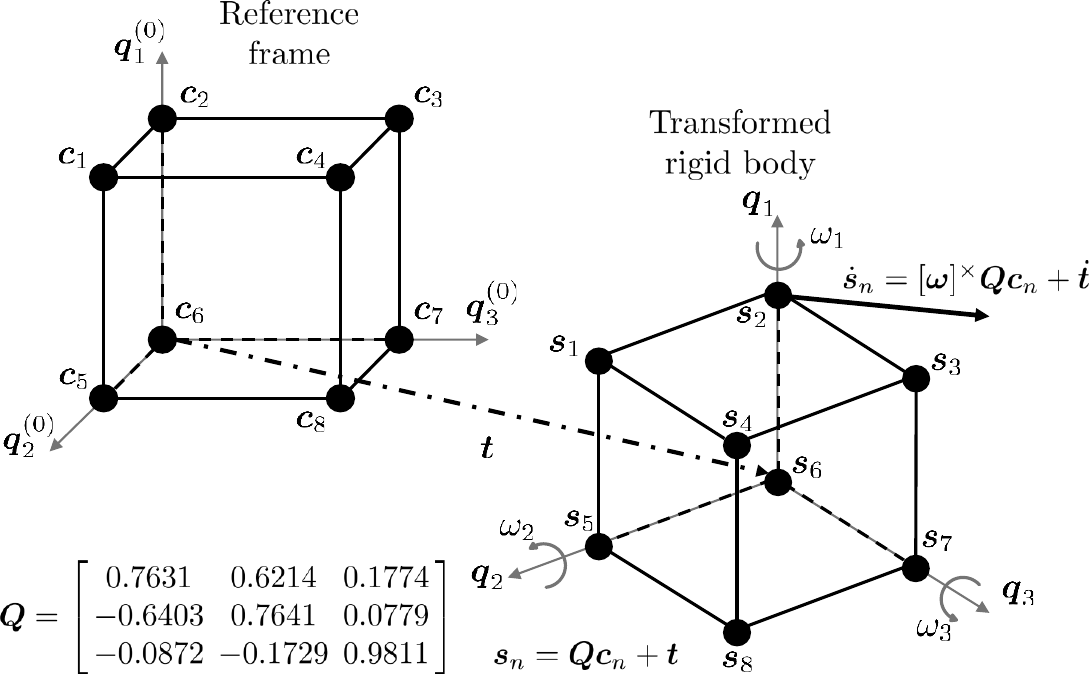}    
\vspace{-2ex}
\caption{An illustration of a moving rigid body, whose location has a \ac{3D} rotation $\boldsymbol{Q}$ and a translation $\boldsymbol{t}$ from the reference frame\protect\footnotemark, as determined by equation \eqref{eq:basic_model_RB}. The velocity $\boldsymbol{\dot{s}}_n$ of the $n$-th sensor as a function of the angular velocity $\boldsymbol{\omega}$ and translational velocity $\boldsymbol{\dot{t}}$, as determined by equation \eqref{eq:basic_model_RB_Mov}, is also illustrated.}
\label{fig:RB_trans_plot}
\vspace{-2ex}
\end{figure}
\noindent where $\bm{Q}_{x}, \,\bm{Q}_{y}, \,\bm{Q}_{z} \in \mathbb{R}^{3\times 3}$ are the roll, pitch, and yaw rotation matrices about the \textit{x-,y-,z-}axes by rotation angles of $\theta_x, \theta_y, \theta_z \in [-180^\circ, 180^\circ]$ degrees, respectively.

In light of the above, the transformed coordinates of the $n$-th sensor after the rotation and translation is described by
\begin{equation}
\label{eq:basic_model_RB}
\boldsymbol{s}_{n} =\boldsymbol{Q} \boldsymbol{c}_{n}+\boldsymbol{t} \in \mathbb{R}^{3 \times 1},
\end{equation}
which is applied identically to all $N$ sensors of the rigid body, as illustrated in Figure \ref{fig:RB_trans_plot}.

The latter system model for stationary rigid bodies can be extended to moving rigid bodies\footnote{{\color{black}It should be noted that the employed model is purely kinematic in nature and does not incorporate additional physical aspects such as the forces acting upon the rigid body or its mass. Furthermore, due to the adopted kinematic formulation, all parameters are considered in a decoupled manner. This simplification can be revisited in a future work, where a coupled model may be adopted and analyzed.}} by introducing the angular velocity vector $\boldsymbol{\omega}\triangleq\left[\omega_{1}, \omega_{2}, \omega_{3}\right]\trans\in\mathbb{R}^{3\times 1}$, and the translational velocity vector $\boldsymbol{\dot{t}}\triangleq [\dot{t}_x, \dot{t}_y, \dot{t}_z]\trans \in\mathbb{R}^{3\times 1}$, 
such that the velocity of the $n$-th sensor can be expressed as

\begin{equation}
\label{eq:basic_model_RB_Mov}
\dot{\boldsymbol{s}}_{n}=[\boldsymbol{\omega}]^{\times} \boldsymbol{Q} \boldsymbol{c}_{n}+\dot{\boldsymbol{t}} \in \mathbb{R}^{3 \times 1},
\end{equation}
where $[\cdot]^{\times}$ is the cross product operator for matrices \cite{Diebel2006RigidBodyAttitude}, which maps the angular velocity vector $\boldsymbol{\omega} $ to a skew-symmetric matrix, given by 
\begin{equation}
\label{eq:omega_matrix}
[\boldsymbol{\omega}]^{\times}=\left[\begin{array}{ccc}
0 & -\omega_{3} & \omega_{2} \\
\omega_{3} & 0 & -\omega_{1} \\
-\omega_{2} & \omega_{1} & 0
\end{array}\right] \in \mathbb{R}^{3 \times 3}.
\end{equation}

\subsection{Sensor Range Measurement Model}

\footnotetext{The reference frame is generally not at the origin, but since a rigid body rotation is only relative to its previous orientation, setting the reference frame to the origin, $i.e.$, making $\boldsymbol{Q}^{(0)} \triangleq \mathbf{I}_{3\times 3}$, can be done \ac{wlg}.}

Consider the problem of performing \ac{RBL} using pairwise range measurements between anchors and sensors, assumed to be available at the anchors, and which are described by
\begin{equation}
\label{eq:range_model}
\tilde{d}_{m,n} = d_{m,n} + w_{m,n} = \left\|\boldsymbol{a}_{m} - \boldsymbol{s}_{n}\right\|_{2} + w_{m,n} \in \mathbb{R},
\end{equation}
with corresponding squared range measurements modeled as
\begin{equation}
\label{eq:sq_range_meas}
\tilde{d}_{m,n}^2 = \left\|\boldsymbol{a}_{m} - \boldsymbol{s}_{n}\right\|_{2}^2 + 2d_{m,n}w_{m,n} + w_{m,n}^2  \in \mathbb{R}.
\end{equation}

In the above, $d_{m,n} \triangleq \left\|\boldsymbol{a}_{m} - \boldsymbol{s}_{n}\right\|_{2}$ is the true Euclidean distance between the $m$-th anchor and the $n$-th sensor, and $w_{m,n} \sim \mathcal{N}(0, \sigma_w^2)$ is the \ac{i.i.d.} \ac{AWGN} of  variance $\sigma_w^2$ affecting the range measurement,
which, following \cite{ChenTSP2015,HoTSP2004,MaICASSP2011}, can be reformulated in terms of a linear relation with a \textit{composite ranging noise} $\xi_{n} \in \mathbb{R}$ given by
\begin{equation}
\label{eq:pos_lin_eq}
\xi_{m,n} = \tilde{d}_{m,n}^{2} - \left\|\boldsymbol{a}_{m}\right\|^{2}_2 -\left\|\boldsymbol{s}_{n}\right\|^{2}_2 +2 \boldsymbol{a}_{m}\trans \boldsymbol{s}_{n} \approx 2d_{m,n}w_{m,n},
\end{equation}
where the second-order noise term $w_{m,n}^2$ is neglected\footnote{\color{black}The second-order noise terms, often neglected in linear approximations, can introduce non-negligible biases in nonlinear estimation problems. Although their influence is typically minor in large scale systems, such as the scenario considered in this work, it may become significant primarily in small scale systems involving nonlinear estimation under high noise conditions.}.

Stacking equation \eqref{eq:pos_lin_eq} for all $M$ anchors and reformulating as a linear system on the $n$-th unknown sensor variable yields \vspace{-3ex}
\begin{align}
\boldsymbol{y}_{n} \!&\triangleq\!\!\!
\begin{bmatrix}
\!\tilde{d}_{1,n}^{2} \!\!-\! \left\|\boldsymbol{a}_{1}\right\|^{2}_2 \!\\
\!\vdots \\
\tilde{d}_{M,n}^{2} \!\!-\! \left\|\boldsymbol{a}_{M}\right\|^{2}_2 \\
\end{bmatrix} 
\!\!\!=\!\!\!
\underbrace{
\begin{bmatrix}
\!-2 \boldsymbol{a}_{1}\trans, &\!\!\!\! 1\; \\
\!\vdots &\!\!\!\! \vdots~\\
\!-2 \boldsymbol{a}_{M}\trans, &\!\!\!\! 1\; \\
\end{bmatrix}}_{\triangleq \,\boldsymbol{G} \,\in\, \mathbb{R}^{M \!\times\!4}}
\!\!\!\!\!\!\!\overbrace{\!
\begin{bmatrix}
\boldsymbol{s}_n \\[1ex]
||\boldsymbol{s}_n||_2^2
\end{bmatrix}\!}^{ ~~~\triangleq \, \boldsymbol{x}_n \, \in \, \mathbb{R}^{4 \!\times\!1}}
\!\!\!\!\!\!+\!\!\!\!\!\!
\underbrace{\!
\begin{bmatrix}
\!\xi_{1,n}\!\\
\!\vdots\!\\
\!\xi_{M,n}\!\\
\end{bmatrix}
\!}_{\triangleq \,\boldsymbol{\xi}_n \,\in\, \mathbb{R}^{M \!\times\!1}}\!\!\!\!\!\!\! \in \!\mathbb{R}^{M \times 1}\!. \nonumber \\[-4ex] 
\label{eq:linear_sys}
\end{align}

In the latter representation of the observed data vector $\boldsymbol{y}_{n} \in \mathbb{R}^{M \times 1}$, {\color{black} the effective channel matrix $\boldsymbol{G}$ $\in \mathbb{R}^{M \times 4}$ has been implicitly defined to be constructed from range measurements and anchor positions $\bm{a}_m$}; the vector $\boldsymbol{x}_n \in \mathbb{R}^{4 \times 1}$ which contains the unknown coordinates of the $n$-sensor, as well as its distance to the origin; and a vector $\boldsymbol{\xi}_{n} \in \mathbb{R}^{M \times 1}$, which gathers the composite noise quantities defined in equation \eqref{eq:pos_lin_eq}.

As hinted in the introduction, the linear system in equation \eqref{eq:linear_sys} can be leveraged for the estimation of the unknown sensor coordinate vector $\boldsymbol{s}_n$ and sensor position norm $||\boldsymbol{s}||_2^2$ in $\boldsymbol{x}_n$, from which equation \eqref{eq:basic_model_RB} can be invoked for the translation and rotation extraction via Procrustes analysis or other classical algorithms \cite{EggertMVA1997}.
We shall return to this idea in Section \ref{sec:proposed_stationary}.

\vspace{-1ex}

\subsection{Sensor Doppler Measurement Model}

In addition to the sensor range measurement model relating to the sensor positions, under the assumption of moving rigid bodies and the corresponding sensors, the Doppler measurement information between the anchors and sensors can also be measured, whose relationship is given by
\begin{equation}
\label{eq:doppler_model}
\tilde{\nu}_{m,n}= \nu_{m,n} + \epsilon_{m,n} = \frac{\left(\boldsymbol{s}_{n}-\boldsymbol{a}_{m}\right)\trans}{d_{m,n}} \dot{\boldsymbol{s}}_{n}+\epsilon_{m,n} \in \mathbb{R},
\end{equation}
where ${\nu}_{m,n} \triangleq \frac{\left(\boldsymbol{s}_{n}-\boldsymbol{a}_{m}\right)\trans}{d_{m,n}} \dot{\boldsymbol{s}}_{n} \in \mathbb{R}$ is the true Doppler shift between $m$-th anchor and $n$-th sensor, and $\epsilon_{m,n} \sim \mathcal{N}(0,  \sigma_\epsilon^2)$ is the \ac{i.i.d.} \ac{AWGN} of the Doppler measurement with noise variance $\sigma_\epsilon^2$.

Multiplying the Doppler measurements of equation \eqref{eq:doppler_model} with the earlier range measurements of equation \eqref{eq:range_model} yields
\begin{equation}
\label{eq:sq_doppler_meas}
\tilde{d}_{m,n} \tilde{\nu}_{m,n} \!=\! \boldsymbol{s}_{n}\trans \dot{\boldsymbol{s}}_{n} \!-\! \boldsymbol{a}_{m}\trans \dot{\boldsymbol{s}}_{n}+{\nu}_{m,n} {w}_{m,n}+{d}_{m,n} \epsilon_{m,n}  \in \mathbb{R},
\end{equation}
which can also be reformulated in terms of a \textit{composite Doppler noise} $\zeta_{{\color{black}m,}n} \in \mathbb{R}$, given by
\begin{equation}
\label{eq:vel_lin_eq}
\zeta_{m, n}\! =\!\tilde{d}_{m,n} \tilde{\nu}_{m,n}\!-\!\boldsymbol{s}_{n}\trans \dot{\boldsymbol{s}}_{n}\!+\! \boldsymbol{a}_{m}\trans \dot{\boldsymbol{s}}_{n} \approx {\nu}_{m,n} {w}_{m,n}\!+\!{d}_{m,n} \epsilon_{m,n}\! \in\! \mathbb{R},
\end{equation}
where the second-order noise term $w_{m,n}^2$ can be considered negligible \cite{ChenTSP2015,HoTSP2004,MaICASSP2011} and is therefore omitted.


Finally, stacking equation \eqref{eq:vel_lin_eq} for all $M$ anchors and reformulating as a linear system on the $n$-th unknown sensor variable yields
\begin{align}
\boldsymbol{\dot{y}}_{n} \!&\triangleq\!\!\!
\begin{bmatrix}
\!\tilde{d}_{1,n} \tilde{\nu}_{1,n} \\
\!\vdots \\
\tilde{d}_{M,n} \tilde{\nu}_{M,n}\\
\end{bmatrix} 
\!\!\!=\!\!\!
\underbrace{
\begin{bmatrix}
\!- \boldsymbol{a}_{1}\trans, &\!\!\!\! 1\; \\
\!\vdots &\!\!\!\! \vdots~\\
\!- \boldsymbol{a}_{M}\trans, &\!\!\!\! 1\; \\
\end{bmatrix}}_{\triangleq \,\boldsymbol{\dot{G}} \,\in\, \mathbb{R}^{M \!\times\!4}}
\!\!\!\!\!\!\!\overbrace{\!
\begin{bmatrix}
\boldsymbol{\dot{s}}_n
\\[1ex]
\boldsymbol{s}_n\trans\boldsymbol{\dot{s}}_n
\end{bmatrix}\!}^{ ~~~\triangleq \, \boldsymbol{\dot{x}}_n \, \in \, \mathbb{R}^{4 \!\times\!1}}
\!\!\!\!\!\!+\!\!\!\!\!\!
\underbrace{\!
\begin{bmatrix}
\!\zeta_{1,n}\!\\
\!\vdots\!\\
\!\zeta_{M,n}\!\\
\end{bmatrix}
\!}_{\triangleq \,\boldsymbol{\zeta}_n \,\in\, \mathbb{R}^{M \!\times\!1}}\!\!\!\!\!\!\! \in \!\mathbb{R}^{M \times 1}\!, \nonumber \\[-4ex] 
\label{eq:linear_sys_mov}
\end{align}
where $\boldsymbol{\dot{y}}_{n} \in \mathbb{R}^{M \times 1}$ and $\boldsymbol{\dot{G}}$ $\in \mathbb{R}^{M \times 4}$ are respectively the observed data vector and effective channel matrix constructed from the measured ranges, Doppler shifts and anchor positions, $\boldsymbol{\dot{x}}_n \in \mathbb{R}^{4 \times 1}$ is the unknown sensor velocity vector, and $\boldsymbol{\zeta}_{n} \in \mathbb{R}^{M \times 1}$ is the vector of composite noise variables, whose elements are given by equation \eqref{eq:vel_lin_eq}.

The linear system in equation \eqref{eq:linear_sys_mov} can be leveraged for the estimation of the unknown sensor velocity vector $\boldsymbol{\dot{s}}_n$ and the multiplication of sensor position and velocity $\boldsymbol{s}_n\trans\boldsymbol{\dot{s}}_n$ in $\boldsymbol{\dot{x}}_n$, from which the unknown rigid body transformation variables can be estimated in a similar fashion to the range measurement-based approach in system.

\section{Proposed Rigid Body Localization}
\label{sec:proposed_stationary}

In this section, we propose a low-complexity position and transformation estimator for \ac{RBL}, in light of the system model derived in Section \ref{sec:system_model} and by leveraging the \ac{GaBP} message passing framework.
Specifically, an initial \ac{GaBP} is derived to solve the position-explicit system in equation \eqref{eq:linear_sys} to obtain {\color{black}jointly the entire 3D} sensor coordinates.
Then, in possession of the initial position estimate, a second \ac{GaBP} is derived on the rigid body transformation parameter-explicit system in equation \eqref{eq:delta_t_lin_syst}, to obtain the final estimate of the \ac{3D} rotation angles $\boldsymbol{\theta}$ and translation vector $\boldsymbol{t}$.

\vspace{-1ex}
\subsection{Stationary Transformation Parameter-based System Model}
\label{sec:reformulation}

In this section, we build upon equation \eqref{eq:linear_sys} towards a reformulation that expresses system variables directly in terms of the \ac{RBL} transformation parameters, \textit{i.e.,} the \ac{3D} rotation angles $\boldsymbol{\theta} \triangleq [\theta_x, \theta_y, \theta_z]\trans \in \mathbb{R}^{3 \times 1}$ and translation vector $\boldsymbol{t}$ \cite{ZhaMRBL2021}.
To that end, we first apply a small-angle approximation\footnote{For practical rigid body tracking applications, subsequent transformation estimations can be assumed to be performed within a sufficiently short time interval such that the change in rotation angle remains small. {\color{black}Although the approximation remains valid for rotation angles up to approximately twenty degrees, convergence of the algorithm can still be achieved for larger angles. However, in such cases, the estimation accuracy may degrade.}} \cite{Diebel2006RigidBodyAttitude} onto the rotation matrix of equation \eqref{eq:rotation_matrix}, obtained by leveraging $\cos\theta \approx 1$ and $\sin\theta \approx \theta$, which yields
\begin{eqnarray}
\label{eq:q_small_angle}
~~~~\bm{Q} \approx\!\!\left[\begin{array}{ccc}
1 & \theta_z & -\theta_y \\
-\theta_z & 1 & \theta_x \\
\theta_y & -\theta_x & 1
\end{array}\right] \in \mathbb{R}^{3 \times 3},
\end{eqnarray}
which in turn can be vectorized into a linear system directly in terms of the Euler angles \cite{ChenTSP2015}, namely
\begin{eqnarray}
\mathrm{vec}(\boldsymbol{Q}) = \boldsymbol{\gamma} + \boldsymbol{L} \boldsymbol{\theta} = \overbrace{
\begin{bmatrix}
1 & 0 & 0 & 0 & 1 & 0 & 0 & 0 & 1
\end{bmatrix}\transs}^{\triangleq\, \boldsymbol{\gamma} \, \in \, \mathbb{R}^{9 \times 1}}&& \nonumber \\
&&
\label{eq:q_vec}
\hspace{-48ex} + \underbrace{
\begin{bmatrix}
0 &  1 & 0 & -1 & 0 & 0 & 0 & 0 & 0 \\ 
0 &  0 & -1 & 0 & 0 & 0 & 1 & 0 & 0 \\ 
0 &  0 & 0 & 0 & 0 & 1 & 0 & -1 & 0 \\ 
\end{bmatrix}\transs\!\!\!}_{\triangleq\, \boldsymbol{L} \, \in \, \mathbb{R}^{9 \times 3}} 
\cdot\!
\begin{bmatrix}
\theta_x \\ \theta_y \\ \theta_z
\end{bmatrix}\!\!.
\end{eqnarray}

Then substituting equation \eqref{eq:q_vec} into equations \eqref{eq:basic_model_RB} and \eqref{eq:pos_lin_eq} and rearranging the terms yields the following alternate representation of the composite noise
\begin{eqnarray}
\xi_{n} \!=&& \hspace{-4ex} \tilde{d}_{m,n}^{2}\!-\!\left\|\boldsymbol{a}_{m}\right\|^{2}_2\! -\! \left\|\boldsymbol{s}_{n}\right\|^{2}_2 \!+\! 2\!\left[\boldsymbol{c}_{n}\trans \otimes \boldsymbol{a}_{m}\trans \right]\!\boldsymbol{\gamma}\!+\!2\!\left[\boldsymbol{c}_{n}\trans \otimes \boldsymbol{a}_{m}\trans\right]\!\boldsymbol{L}\boldsymbol{\theta}\nonumber \\
&&\hspace{-4ex} +2 \boldsymbol{a}_{m}\trans \boldsymbol{t}, \in \mathbb{R},\!\! \label{eq:delta_lin_eq}
\end{eqnarray}
where the matrix product vectorization identity $\mathrm{vec}(\mathbf{X Y Z}) = (\mathbf{Z}\trans \otimes \mathbf{X}) \mathrm{vec}(\mathbf{Y})$ has been used, with $\otimes$ denoting the Kronecker product operator.

In light of the above, the fundamental system can be rewritten leveraging the linearization of equation \eqref{eq:delta_lin_eq},
\begin{subequations}
\label{eq:delta_t_lin_syst}
\begin{equation}
\boldsymbol{z}_{n} = \boldsymbol{H}_{\theta} \!\cdot\! \boldsymbol{\theta} + \boldsymbol{H}_{t} \!\cdot\! \boldsymbol{t} + \boldsymbol{\xi}_{n} \in \mathbb{R}^{M \times 1},
\end{equation}
with
\begin{equation}
\boldsymbol{z}_{n} \!=\!\! \left[\!\!\begin{array}{c}
\tilde{d}_{1,n}^{2}-\left\|\boldsymbol{a}_{1}\right\|^{2}_2- \left\|\boldsymbol{s}_{n}\right\|^{2}_2 + 2\left[\boldsymbol{c}_{n}\trans \otimes \boldsymbol{a}_{1}\trans \right]\!\boldsymbol{\gamma} \\[1ex]
\vdots \\[1ex]
\tilde{d}_{M,n}^{2}-\left\|\boldsymbol{a}_{M}\right\|^{2}_2 - \left\|\boldsymbol{s}_{n}\right\|^{2}_2 + 2\left[\boldsymbol{c}_{n}\trans \otimes \boldsymbol{a}_{M}\trans \right]\!\boldsymbol{\gamma}
\end{array}\!\!\right] \!\! \in \mathbb{R}^{M \times 1},
\end{equation}
where $\boldsymbol{z}_{n} \in \mathbb{R}^{M \times 1}$ is the effective observed data vector, $\boldsymbol{\xi}_{n} \in \mathbb{R}^{M \times 1}$ is the vector of composite noise variables from equation \eqref{eq:pos_lin_eq}, and {\color{black}the two effective channel matrices for rotation and translation respectively defined as}

\begin{align}
  \color{black}
\boldsymbol{H}_{\theta} \!=\!\! \left[\begin{array}{c}
\!\!\!\!\!-2\!\left[\boldsymbol{c}_{i}\trans \otimes \boldsymbol{a}_{1}\trans \right]\!\boldsymbol{L}\!\!\!\\
\vdots   \\[0.5ex]
\!\!\!\!-2\!\left[\boldsymbol{c}_{i}\trans \otimes \boldsymbol{a}_{M}\trans \right]\!\boldsymbol{L}\!\!\!
\end{array}\right] \!\! \in \!\mathbb{R}^{M \times 3}\!, 
~ \boldsymbol{H}_{t} \!=\!\! \left[\begin{array}{c}
\!\!\!\!-2 \boldsymbol{a}_{1}\trans\!\!\!\!  \\
\vdots \\
\!\!\!\!-2 \boldsymbol{a}_{M}\trans\!\!\! 
\end{array}\right]\!\! \in \!\mathbb{R}^{M \times 3}.
\end{align}
\end{subequations}

%


\subsection{Linear GaBP for Sensor Position Estimation}
\label{sec:pos_est_sol}

In what follows, we will derive the step-by-step \ac{GaBP} message-passing rules for \ac{RBL} based on the linear model of equation \eqref{eq:linear_sys}.
Since the derivation is identical for each $n$-th sensor node, we shall for the sake of convenience temporarily drop the subscript $_n$ in the quantities $\boldsymbol{y}_{n}$, $\boldsymbol{x}_{n}$ and $\boldsymbol{\xi}_n$ in equation \eqref{eq:linear_sys}, such that we can define the elements of each of these vectors respectively as $\boldsymbol{y}_{n} \to \boldsymbol{y} = \left[y_{1}, \ldots, y_{m}, \ldots, y_{M}\right]\trans, \in \mathbb{R}^{M \times 1}$, $\boldsymbol{x}_{n} \to \boldsymbol{x} = \left[x_{1}, \ldots, x_{k}, \ldots, x_{K+1}\right]\trans, \in \mathbb{R}^{K+1 \times 1}$, with $k=\{1,\ldots,K\!+\!1\}$ where $K$ is the dimension of the space, and $\boldsymbol{\xi}_{n} \to \boldsymbol{\xi} = \left[\xi_{1}, \ldots, \xi_{m}, \ldots, \xi_{M}\right]\trans, \in \mathbb{R}^{M \times 1}$.
In addition, let $g_{m,k}$ denote the entries of the matrix $\boldsymbol{G}$.

\begin{figure}[H]
\centering
{\includegraphics[width=1\columnwidth]{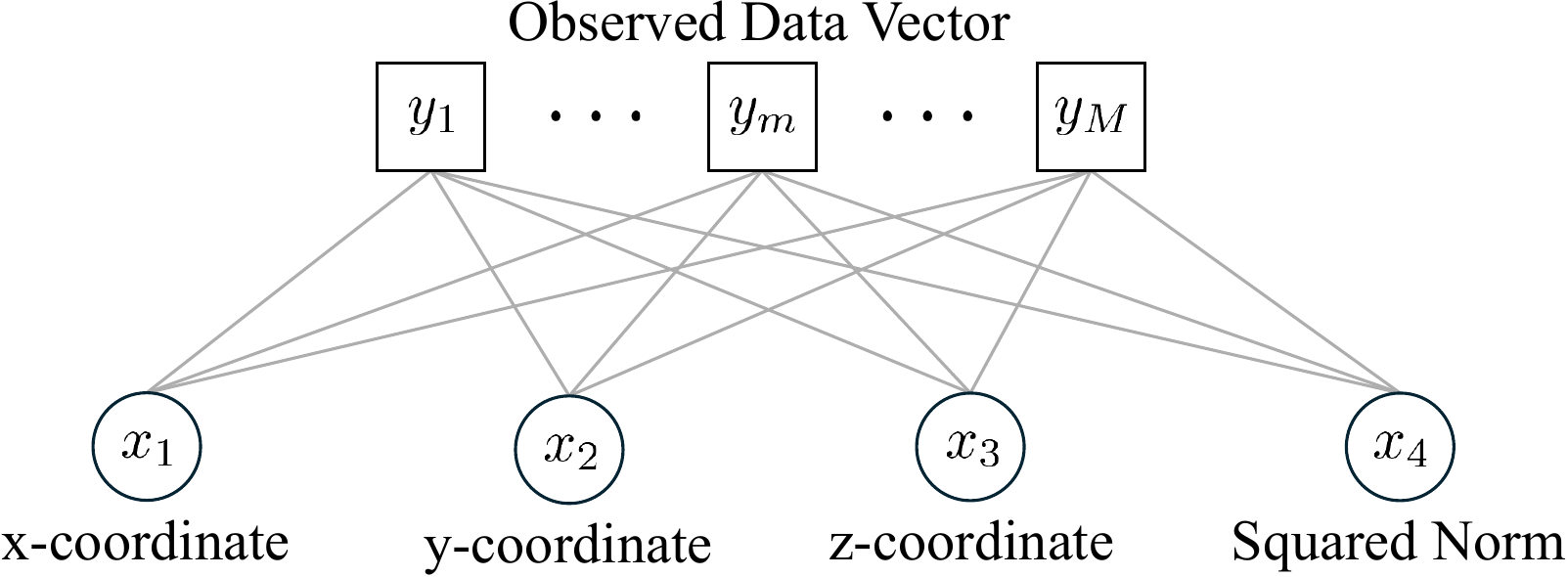}} 
\caption{\color{black}Factor graph structure of the linear system in equation \eqref{eq:linear_sys}, where the variable nodes (circles) represent the unknown variables $x_k$ and the factor nodes (squares) represent the observations $y_m$. Edges are defined by the connecting entries of the effective channel matrix $\boldsymbol{G}$.}
\label{fig:factorgraph_y}
\vspace{-1ex}
\end{figure}

{\color{black}

The proposed \ac{GaBP} algorithm operates on a linear system of the form $\boldsymbol{y} = \boldsymbol{G}\boldsymbol{x} + \boldsymbol{\xi}$, where $\boldsymbol{x}$ denotes the collection of unknown variables, such as sensor positions, velocities, or rigid body parameters as discussed in the following sections. The matrix $\boldsymbol{G}$ represents the system matrix determined by the geometric configuration of anchors and sensors.

The associated factor graph is bipartite, comprising variable nodes corresponding to each element of $\boldsymbol{x}$ and factor nodes for each observation element of $\boldsymbol{y}$, with connections defined by the nonzero entries in $\boldsymbol{G}$, as illustrated in Figure \ref{fig:factorgraph_y}.
As evident from the system formulation in the previous section, this graph structure is sparse in practical deployments and the resulting system is well conditioned\footnote{\color{black}In all scenarios considered in the following sections, the system matrices are well conditioned given appropriate physical layouts of sensors and anchors. This characteristic is known to influence the convergence of \ac{GaBP}, subject to conditions such as walk summability or diagonal dominance of the information matrix. As detailed in Section V, where a comprehensive performance analysis of the proposed \ac{GaBP} algorithms is provided, empirical evaluations have demonstrated consistent convergence across all tested configurations.} under suitable arrangements.}

Under such simplified notation, an estimate of the $k$-th element $x_k$ of the position vector $\boldsymbol{x}$ of a given $n$-th sensor node, from the $m$-th observation $y_m$ of the received signal $\boldsymbol{y}$ is given by a soft replica denoted by $\hat{x}_{m,k}^{[j]}$, whose \ac{MSE} is given by
\begin{equation}
\label{eq:pos_MSE}
\psi_{m,k}^{[j]} = \mathbb{E}_{\mathsf{x}_{k}}\!\!\left\{\big|x_{k}-\hat{x}_{m, k}^{[j]}\big|^{2}\right\}=\mathbb{E}_{\mathsf{x}_{k}}\!\!\left\{x_{k}^{2}\right\}-\Big(\hat{x}_{m, k}^{[j]}\Big)^{\!2} \in \mathbb{R},
\end{equation}
where the superscript $(\,\cdot\,)^{[j]}$ denotes a quantity at $j$-th iteration of the \ac{GaBP}.

\newpage
The first step of \ac{GaBP} is the \ac{sIC} on the $m$-th element of the observed data $y_{m}$ for the estimation of the $k$-th estimated variable, described by
\begin{equation}
\label{eq:pos_soft_IC}
\begin{aligned}
\tilde{y}_{m,k}^{[j]} &=y_{m}-\sum_{i \neq k} g_{m, i} \hat{x}_{m, i}^{[j]} \\
&=g_{m, k} x_{k}+\underbrace{\sum_{i \neq k} g_{m, i}\big(x_{i}-\hat{x}_{m, i}^{[j]}\big)+\xi_{m}}_{\triangleq \, \alpha_{m,k}^{[j]} \,  \in \, \mathbb{R}} \in \mathbb{R},
\end{aligned}
\end{equation}
where $\tilde{y}_{m,k}^{[j]} \in \mathbb{R}$ is the \ac{sIC} symbol corresponding to the $m$-th element of the observed data vector and the $k$-th element of the position vector, and $g_{m,k} \in \mathbb{R}$ is the $(m,k)$-th element of the effective channel matrix $\boldsymbol{G}$ in equation {\color{black}\eqref{eq:linear_sys}}.

Assuming that the interference-plus-noise term $\alpha_{m,k}^{[j]} \in \mathbb{R}$ in equation \eqref{eq:pos_soft_IC} follows a normal distribution under the scalar Gaussian approximation \cite{Chockalingam_Rajan_2014}, the conditional \ac{PDF} of $\tilde{y}_{x:m,k}^{[j]}$ can be expressed as
\begin{equation}
p_{\tilde{\mathsf{y}}_{m,k}^{[j]} \mid \mathsf{x}_{k}} \!\Big(\tilde{y}_{m,k}^{[j]} \big| x_{k}\Big) \propto \mathrm{exp} \bigg[-\frac{\big|\tilde{y}_{m,k}^{[j]}-g_{m, k} x_{k}\big|^{2}}{ \sigma_{m,k}^{2[j]}}\bigg]\!, \label{eq:pos_cond_PDF}
\end{equation}
whose conditional variance is given by
\begin{equation}
\label{eq:pos_var}
\sigma_{m,k}^{2[j]}=\sum_{i \neq k} |g_{m, i}|^{2} \psi_{m, i}^{[j]} + N_{0} \in \mathbb{R},
\end{equation}
where $N_{0}$ is the power of the composite noise quantity described in equation \eqref{eq:delta_lin_eq}.

In hand of the conditional \acp{PDF} of the \ac{sIC} symbols, the extrinsic \ac{PDF} of $x_{k}$ can be obtained via
\begin{equation}
\label{eq:pos_extr_PDF}
\prod_{i \neq m} p_{\tilde{\mathsf{y}}_{i,k}^{[j]} \mid \mathsf{x}_{k}} \!\!\left(\tilde{y}_{i,k}^{[j]} \mid x_{k}\right) \propto \mathrm{exp}
\! \bigg[-\frac{\big|x_{k}-\bar{x}_{m,k}^{[j]}\big|^{2}}{\bar{v}_{m,k}^{[j]}}\bigg],
\end{equation}
with the extrinsic mean and variance respectively given by
\begin{subequations}
\label{eq:pos_extr}
\begin{equation}
\label{eq:pos_extr_mean}
\bar{x}_{m,k}^{[j]} =\bar{v}_{m,k}^{[j]} \cdot \sum_{i \neq m} \frac{g_{i, k} \!\cdot\! \tilde{y}_{i,k}^{[j]}}{\sigma_{i,k}^{2[j]}}  \in \mathbb{R},
\end{equation}
\begin{equation}
\label{eq:pos_extr_var}
\bar{v}_{m,k}^{[j]} =\bigg(\sum_{i \neq m} \frac{|g_{i, k}|^{2}}{\sigma_{i,k}^{2[j]}}\bigg)^{\!\!-1} \!\!\!\!\!\in \mathbb{R}.
\end{equation}
\end{subequations}

{\color{black}
The extrinsic mean $\bar{x}_{m,k}^{[j]}$ and variance $\bar{v}_{m,k}^{[j]}$ are subsequently combined with the prior distribution\footnote{\color{black}The statistics of the prior distribution of the \ac{RBL} parameters are assumed to be known, which is a common assumption in the literature \cite{ChenTSP2015, Jiang2019}, but can also be learned from the data within the iterations of the \ac{GaBP} via methods such as expectation-maximization approaches \cite{Ranasinghe2025JCDE} if they are assumed to be not available a priori, where the latter is out of scope of this article.} of the \ac{RBL} position variables to obtain the denoised posterior estimates.

In accordance with the \ac{GaBP} framework, the Bayes-optimal denoiser is derived from the product of the extrinsic \ac{PDF} of the position variable and the prior distribution, which are both assumed to be Gaussian\footnote{If an {\color{black}alternative} prior distribution of the position variables are assumed, \textit{i.e.,} uniform distribution, a different Bayes-optimal denoiser can be utilized.}, such that the posterior distribution is Gaussian with mean $\check{x}_{m, k}^{[j]}$ and variance $\check{\psi}_{m,k}^{[j]}$ given by
\begin{equation}
    \check{x}_{m, k}^{[j]} = \frac{\bar{v}_{m,k}^{[j]} \mu_x + \phi_x \bar{x}_{m,k}^{[j]}}{\phi_x + \bar{v}_{m,k}^{[j]}} \text{ and }  \check{\psi}_{m,k}^{[j]} = \frac{\phi_x \bar{v}_{m,k}^{[j]}}{\phi_x+\bar{v}_{m,k}^{[j]}} 
\end{equation}

\begin{algorithm}[H]
\caption{: Linear \ac{GaBP} for Position Estimation}
\label{alg:pos_GaBP}
\hspace*{\algorithmicindent}
\begin{algorithmic}[1]
\vspace{-0.9ex}
\Statex \hspace{-4ex} \textbf{Input:} $\boldsymbol{y}_n \!~\forall n, \; \boldsymbol{G}, \;\phi_x, \;N_0, \;j_\mathrm{max}, \;\rho$. \vspace{-1.25ex}
\Statex \hspace{-4.4ex} \hrulefill
\Statex \hspace{-4ex}  \textbf{Output:} $\tilde{x}_{k} \!~\forall k$ (for all sensor nodes $\forall n$); \vspace{-1.25ex}
\Statex \hspace{-4.4ex} \hrulefill
\Statex \hspace{-3.2ex} \textit{{Perform}} $\forall n, m, k:$ \vspace{0.25ex}
\State Initialize $\hat{x}_{m, k}^{[1]}$ and $\psi_{m, k}^{[1]}$;
\For {$j = 1$ to $j_\mathrm{max}$}
\State \hspace{-1.5ex} Compute the \ac{sIC} symbol $\tilde{y}_{m,k}^{[j]}$ via eq. \eqref{eq:pos_soft_IC};
\State \hspace{-1.5ex} Compute the conditional variance $\sigma_{m,k}^{2[j]}$ via eq. \eqref{eq:pos_var};
\State \hspace{-1.5ex} Compute the extrinsic mean $\bar{x}_{m,k}^{[j]}$ via eq. \eqref{eq:pos_extr_mean};
\State \hspace{-1.5ex} Compute the extrinsic variance $\bar{v}_{m,k}^{[j]}$ via eq. \eqref{eq:pos_extr_var};
\State \hspace{-1.5ex} Denoise the beliefs $\check{x}_{m, k}^{[j]}, \check{\psi}_{m, k}^{[j]}$ via eq. \eqref{eq:pos_denoised};
\State \hspace{-1.5ex} Update the soft-replicas with damping via eq. \eqref{eq:damped_update}; \vphantom{ $\check{x}_{m, k}^{[j]}$}
\EndFor
\State Obtain final consensus estimate $\tilde{x}_{k}$ via eq. \eqref{eq:pos_final_est}; 
\vspace{-2ex}
\end{algorithmic} 
\hspace*{\algorithmicindent}
\end{algorithm}

\noindent with $\mu_x \in \mathbb{R}$ and $\phi_x \in \mathbb{R}$ being the mean and variance of the prior distribution of the position variable, where since it is assumed $\mu_x = 0$, the final denoised posterior estimates are simplified to
\begin{equation}
\label{eq:pos_denoised}
\check{x}_{m, k}^{[j]}=\frac{\phi_x \bar{x}_{m, k}^{[j]}}{\phi_x +\bar{v}_{m,k}^{[j]}} \in \mathbb{R} ~\,\text{and}\,~
\check{\psi}_{m,k}^{[j]}=\frac{\phi_x \bar{v}_{m,k}^{[j]}}{\phi_x+\bar{v}_{m,k}^{[j]}}\in \mathbb{R}. \!\!\!
\end{equation}
}

Finally, the $(j\!+\!1)$-th soft-replica is obtained via the following damped update to prevent error floors caused by early erroneous convergence to a local optima
\begin{subequations}
\begin{align}
\hat{x}_{m, k}^{[j+1]} = \rho \hat{x}_{m, k}^{[j]} + (1 - \rho) \check{x}_{m, k}^{[j]}, \\[1ex]
\psi_{m, k}^{[j+1]} = \rho {\psi}_{m, k}^{[j]} + (1 - \rho) \check{\psi}_{m, k}^{[j]},
\end{align}
\label{eq:damped_update}
\end{subequations}
where $\rho \in [0,1]$ is a selected damping parameter.

At the end of the \ac{GaBP} iterations, with the last iteration denoted by $j_\mathrm{max}$, the final estimate of the position variable is obtained via a consensus belief combination, given by
\begin{equation}
\label{eq:pos_final_est}
\tilde{x}_{k} = \left( \sum_{m = 1}^{M} \frac{|g_{m, k}|^2}{\sigma_{m,k}^{2[j_\mathrm{max}]}} \right)^{\!\!\!-1} \!\!\!\left( \sum_{m = 1}^{M} \frac{g_{m, k} \cdot \tilde{y}_{m,k}^{[j_\mathrm{max}]}}{ \sigma_{m,k}^{2[j_\mathrm{max}]}} \right) \in \mathbb{R}.
\end{equation}


Algorithm \ref{alg:pos_GaBP} summarizes the proposed method for the sensor position estimation without rigid body conformation, based on the range information between the sensors and anchors, ultimately yielding $(K+1)$ elements  in $\boldsymbol{x}_n$ composed of $K$ coordinate estimates of $\mathbf{s}_n$ and its absolute norm $||\boldsymbol{s}_n||_2^2$.

Subsequently, the estimated position information (norm of the sensor coordinate) can be used to construct the second linear equation of equation \eqref{eq:delta_t_lin_syst} incorporating the rigid body conformation in terms of the rotation angles and translation vector, as will be leveraged in the following subsection to derive the second \ac{GaBP} algorithm to directly estimate the rigid body transformation parameters.


\subsection{Bivariate GaBP for Transformation Parameter Estimation}
\label{sec:del_t_est_sol}

\begin{figure}[H]
\centering
{\includegraphics[width=1\columnwidth]{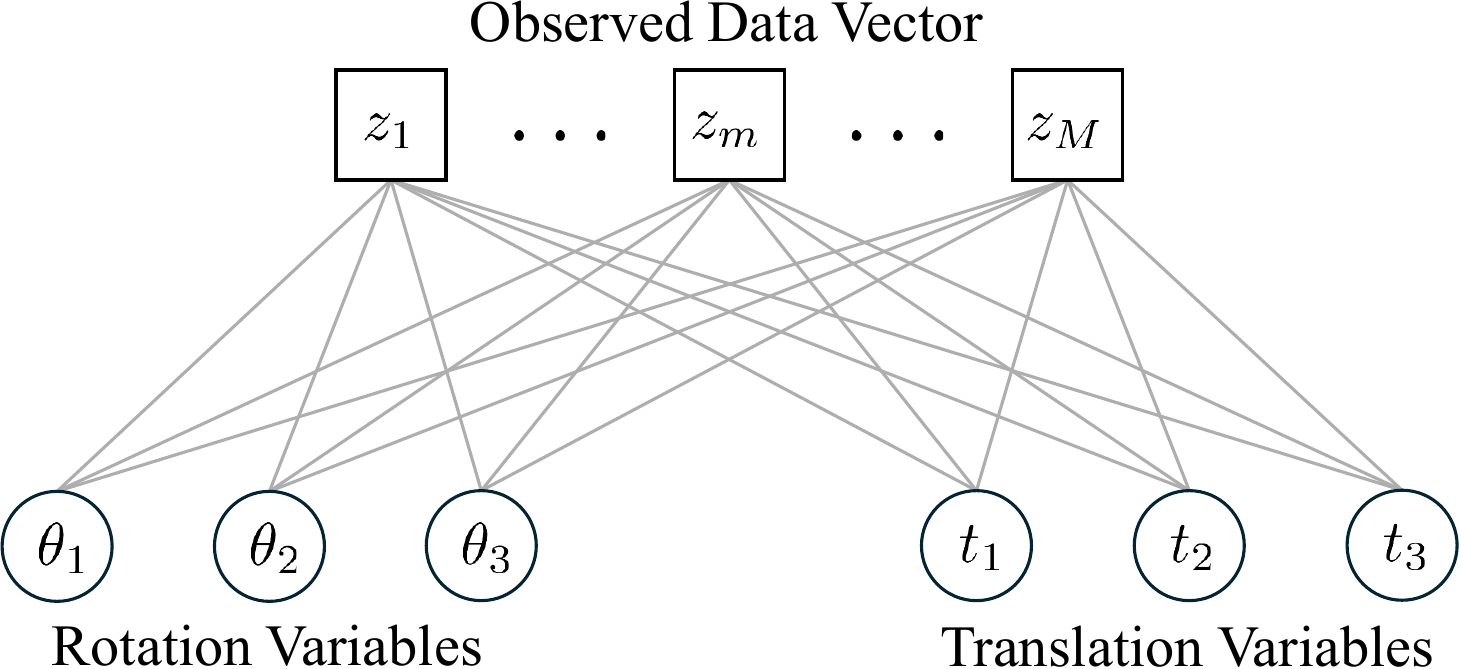}} 
\caption{\color{black}Factor graph structure of the bivariate system in equation \eqref{eq:delta_t_lin_syst}, where the variable nodes (circles) represent the unknown variables $\theta_k$ and $t_\ell$ and the factor nodes (squares) represent the transformed effective observations $z_m$. Edges are defined by the connecting entries of the effective channel matrices $\boldsymbol{H}_\theta$ and $\boldsymbol{H}_t$.}
\label{fig:factorgraph_biv}
\vspace{-5ex}
\end{figure}

While the linear formulation and the message passing rules for the \ac{GaBP} iterations are very similar to Algorithm \ref{alg:pos_GaBP}, in the case of transformation parameter estimation based on equation \eqref{eq:delta_t_lin_syst}, there exist two sets of variables $\theta_k$ with $k \in \{1,\ldots,K\}$ and $t_\ell$ with $\ell \in \{1,\ldots,K\}$ {\color{black}as illustrated by the factor graph structure of Figure \ref{fig:factorgraph_biv}} such that the \ac{GaBP} rules are elaborated separately.
To elaborate, first \ac{sIC} is performed on the observed information respectively for the angle and translation variables as
\begin{subequations}
\label{eq:del_t_soft_IC}
\begin{align}\label{eq:del_IC}
\tilde{z}_{\theta:m,k}^{[j]} &= z_{m} - \sum_{i \neq k} h_{\theta:m,i}\hat{\theta}_{m,i}^{[j]} - \sum_{i = 1}^{K} h_{t:m,i}\hat{t}_{m,i}^{[j]}, \\
&= h_{\theta:m,k}\theta_{k} +  \sum_{i = 1}^{K} h_{t:m,i}(t_{i} - \hat{t}_{m,i}^{[j]}) \nonumber \\[-1ex]
& \hspace{17ex} + \sum_{i \neq k} h_{\theta:m,i}(\theta_{i} - \hat{\theta}_{m,i}^{[j]}) + \xi_m, \nonumber
\end{align}
\vspace{-3ex}
\begin{align}
\label{eq:t_IC}
\tilde{z}_{t:m,\ell}^{[j]} &= z_{m} - \sum_{i = 1}^{K} h_{\theta:m,i}\hat{\theta}_{m,i}^{[j]} - \sum_{i \neq \ell}  h_{t:m,i}\hat{t}_{m,i}^{[j]}, \\
&=  h_{t:m,\ell}t_{\ell} + \sum_{i = 1}^{K}  h_{\theta:m,i}(\theta_{i} - \hat{\theta}_{m,i}^{[j]}) \nonumber \\[-1ex] 
& \hspace{17ex} + \sum_{i \neq \ell}  h_{t:m,i}(t_{i} - \hat{t}_{m,i}^{[j]}) + \xi_m. \nonumber
\end{align}
\end{subequations}

In turn, the conditional \acp{PDF} of the \ac{sIC} symbols are given by
\begin{equation}
\label{eq:del_t_cond_PDF}
\begin{aligned}
p_{\tilde{\mathrm{z}}_{\theta:m,k}^{[j]} \mid \mathrm{\theta}_{k}}(\tilde{z}_{\theta:m,k}^{[j]}|\theta_{k}) &\propto \mathrm{exp}\bigg[ -\frac{|\tilde{z}_{\theta:m,k}^{[j]} - h_{\theta:m,k} \theta_{k}|^2}{\sigma_{\theta:m,k}^{2\,[j]}} \bigg], \\
p_{\tilde{\mathrm{z}}_{t:m,\ell}^{[j]} \mid \mathrm{t}_{\ell}}(\tilde{z}_{t:m,\ell}^{[j]}|t_{\ell}) &\propto \mathrm{exp}\bigg[ -\frac{|\tilde{z}_{t:m,\ell}^{[j]} - h_{t:m,\ell} t_{\ell}|^2}{\sigma_{t:m,\ell}^{2\,[j]}} \bigg],
\end{aligned}
\end{equation}
with conditional variances
\begin{subequations}
\label{eq:del_t_theta_var}
\begin{equation}
\label{eq:del_theta_var}
\sigma_{\theta:m,k}^{2\,[j]}\! = \!\displaystyle\sum\limits_{i \neq k} \big|h_{\theta:m,i}\big|^2\psi_{\theta:m,i}^{[j]}\! +\! \sum\limits_{i = 1}^{K} \big|h_{t:m,i}\big|^2\psi_{t:m,i}^{[j]} \!+\! N_{0} \in \mathbb{R},
\end{equation}
\begin{equation}
\label{eq:del_t_var}
\sigma_{t:m,\ell}^{2\,[j]} \!=\! \displaystyle\sum\limits_{i = 1}^{K} \big|h_{\theta:m,i}\big|^2\psi_{\theta:m,i}^{[j]}\! +\! \sum\limits_{i \neq \ell} \big|h_{t:m,i}\big|^2\psi_{t:m,i}^{[j]}\! +\! N_{0} \in \mathbb{R},
\end{equation}
with the corresponding \acp{MSE} given by  $\psi_{\theta:m,k}^{[j]} = \mathbb{E}_{\mathsf{\theta}_k}\!\big[ | \theta_{k} - \hat{\theta}_{m,k}^{[j]} |^2 \big]$ and $\psi_{t:m,\ell}^{[j]} = \mathbb{E}_{\mathsf{t}_{\ell}}\!\big[ | t_{\ell} - \hat{t}_{m,\ell}^{[j]} |^2 \big]$, respectively.
\end{subequations}

With the conditional \acp{PDF} in hand, the extrinsic \ac{PDF} is obtained as \vspace{-1.5ex}
\begin{equation}
\begin{aligned}
\prod_{i \neq m} p_{\tilde{\mathsf{z}}_{\theta:i,k}^{[j]} \mid \mathsf{\theta}_{k}}\left(\tilde{z}_{\theta:i,k}^{[j]} \mid \theta_{k}\right) &\propto \mathrm{exp}\bigg[ -\frac{|\theta_{k} - \bar{\theta}_{m,k}^{[j]}|^2}{\bar{v}_{\theta:m,k}^{[j]}} \bigg], \\
\prod_{i \neq m} p_{\tilde{\mathsf{z}}_{t:i,\ell}^{[j]} \mid \mathrm{t}_{\ell}}\left(\tilde{z}_{t:i,\ell}^{[j]} \mid t_{\ell}\right) &\propto \mathrm{exp}\bigg[ -\frac{|t_{\ell} - \bar{t}_{m,\ell}^{[j]}|^2}{\bar{v}_{t:m,\ell}^{[j]}} \bigg],
\end{aligned}
\label{eq:del_t_extr_PDF}
\end{equation}
where the extrinsic means and variances are given by
\begin{subequations}
\label{eq:del_t_theta_extr_mean}
\begin{align}
\bar{\theta}_{m,k}^{[j]} &= \bar{v}_{\theta:m,k}^{[j]} \bigg( \sum_{i \neq m} \frac{h_{\theta:i,k} \cdot \tilde{z}_{\theta:i,k}^{[j]}}{ \big(\sigma_{\theta:i,k}^{[j]}\big)^2} \bigg)\in \mathbb{R}, \label{eq:del_the_extr_mean}\\[1ex]
\bar{t}_{m,\ell}^{[j]} &= \bar{v}_{t:m,\ell}^{[j]} \bigg( \sum_{i \neq m} \frac{h_{t:i,\ell} \cdot \tilde{z}_{t:i,\ell}^{[j]}}{ \big(\sigma_{t:i,\ell}^{[j]}\big)^2} \bigg)\in \mathbb{R}, \label{eq:del_t_extr_mean}
\end{align}
\end{subequations}

\begin{subequations}
\label{eq:del_t_theta_extr_var}
\begin{align}
\bar{v}_{\theta:m,k}^{[j]} &= \bigg( \sum_{i \neq m} \frac{|h_{\theta:i,k}|^2}{\big(\sigma_{\theta:i,k}^{[j]}\big)^2} \bigg)^{\!\!\!-1}  \!\!\!\!\in \mathbb{R}, \label{eq:del_the_extr_var} \\[1ex]
\bar{v}_{t:m,\ell}^{[j]} &= \bigg( \sum_{i \neq m} \frac{|h_{t:i,\ell}|^2}{\sigma_{t:m,\ell}^{2\,[j]}} \bigg)^{\!\!\!-1} \!\!\!\!\in \mathbb{R}. \label{eq:del_t_extr_var}
\end{align}
\end{subequations}

Finally{\color{black},} the denoisers with a Gaussian prior are given by 
\begin{subequations}
\begin{equation}
\label{eq:del_t_soft_est}
\begin{aligned}
\check{\theta}_{m,k} &= \frac{\phi_{\theta} \cdot \bar{\theta}_{m,k}^{[j]}}{\phi_{\theta} + \bar{v}_{\theta:m,k}^{[j]}} \in \mathbb{R}, 
& \check{t}_{m,\ell} &= \frac{\phi_{t} \cdot \bar{t}_{m,\ell}^{[j]}}{\phi_{t} + \bar{v}_{t:m,\ell}^{[j]}} \in \mathbb{R}, \\
\end{aligned}
\end{equation}
\begin{equation}
\label{eq:del_t_est_mse}
\hspace{-2ex}\check{\psi}_{\theta:m,k} = \frac{\phi_{\theta} \cdot \bar{v}_{\theta:m,k}^{[j]}}{\phi_{\theta}\! +\! \bar{v}_{\theta:m,k}^{[j]}} \in \mathbb{R},\;
 \check{\psi}_{t:m,\ell} = \frac{\phi_{t} \cdot \bar{v}_{t:m,\ell}^{[j]}}{\phi_{t}\! +\! \bar{v}_{t:m,\ell}^{[j]}} \in \mathbb{R}, 
\end{equation}
where $\phi_{\theta}$ and $\phi_{t}$ are the variance of the elements in $\boldsymbol{\theta}$ and $\boldsymbol{t}$.
\end{subequations}


Subsequently, the soft-replicas are iteratively updated similarly to equation \eqref{eq:damped_update}, with a damping factor $\rho$, for  $j_\mathrm{max}$ iterations of the message passing algorithm or {\color{black}until a convergence criteria is met}, after which the consensus estimates are obtained as 
\begin{subequations}
\label{eq:del_t_final_est}
\begin{eqnarray}
&\tilde{\theta}_{k} = \bigg( \sum\limits_{m = 1}^{M} \frac{|h_{\theta:m,k}|^2}{\big(\sigma_{\theta:m,k}^{[j_\mathrm{max}]}\big)^2} \bigg)^{\!\!\!-1} \! \! \bigg( \sum\limits_{m = 1}^{M} \frac{h_{\theta:m,k} \cdot \tilde{z}_{\theta:m,k}^{[j_\mathrm{max}]}}{ \big(\sigma_{\theta:m,k}^{[j_\mathrm{max}]}\big)^2} \bigg) \in \mathbb{R},\;\;\;& \label{eq:del_t_final_est_theta} \\
&\tilde{t}_{\ell} = \bigg( \sum\limits_{m = 1}^{M} \frac{|h_{t:m,\ell}|^2}{\big(\sigma_{t:m,\ell}^{[j_\mathrm{max}]}\big)^2} \bigg)^{\!\!\!-1} \!\bigg( \sum\limits_{m = 1}^{M} \frac{h_{t:m,\ell} \cdot \tilde{z}_{t:m,\ell}^{[j_\mathrm{max}]}}{\big(\sigma_{t:m,\ell}^{[j_\mathrm{max}]}\big)^2} \,\bigg) \in \mathbb{R}.\;\;\;&
\label{eq:del_t_final_est_t}
\end{eqnarray}
\end{subequations}

While the message passing rules elaborated by equations \eqref{eq:del_t_soft_IC}-\eqref{eq:del_t_final_est} are complete to yield the estimated rotation angles and {\color{black}translation} vectors{\color{black}, there is a difference in effective channel powers of $\boldsymbol{H}_{\theta}$ and $\boldsymbol{H}_{t}$ in equation \eqref{eq:delta_t_lin_syst}, where the latter is typically much larger, due to the absolute positions of the anchors and sensors}\footnotemark.
Such significant {\color{black}differences} in effective channel powers lead to good estimation performance of the translation vector elements, but erroneous estimation performance of the rotation angles in a joint estimation described by the \ac{GaBP} procedure.

This behavior can also be intuitively understood by considering the illustration in Figure \ref{fig:RB_trans_plot}, where a small rotation of the rigid body is expected to have a less prominent effect on the absolute sensor positions then the translation, as assumed in the system formulation of Section \ref{sec:system_model}.
\newpage

\footnotetext{The effective channel powers are highly {\color{black}dependent} on the sensor and anchor deployment structure, and for typical indoor sensing scenarios as illustrated in Figure \ref{fig:sys_mod_plot}, the anchor coordinates are of larger absolute value than the rigid body sensor coordinates, leading to the large power difference.}

\begin{algorithm}[H]
\caption{: Double \ac{GaBP} for \ac{RBL} Parameter Estimation}
\label{alg:RBL_GaBP}
\hspace*{\algorithmicindent}
\begin{algorithmic}[1]
\vspace{-0.9ex}
\Statex \hspace{-4ex} \textbf{Input:} $\boldsymbol{z}_n \,(||\boldsymbol{s}_n||_2^2) \!~\forall n,  \boldsymbol{H}_{\theta}, \boldsymbol{H}_{t}, \phi_{\theta}, \phi_{t}, N_0, j_\mathrm{max}, \rho$. \vspace{-1.25ex}
\Statex \hspace{-4.4ex} \hrulefill
\Statex \hspace{-4ex}  \textbf{Output:} $\tilde{\theta}_{k}$ and $\tilde{t}_{\ell} ~\forall k,\ell$ (for all sensor nodes $\forall n$); \vspace{-1.5ex}
\Statex \hspace{-4.4ex} \hrulefill 
\Statex \hspace{-3.2ex} \textit{{Perform}} $\forall n, m, k, \ell:$ \vspace{0.25ex}
\State {\color{black}Initialize} $\hat{\theta}_{m, k}^{[1]}$,  $\hat{t}_{m, \ell}^{[1]}$, $\psi_{\theta:m, k}^{[1]}$, $\psi_{t:m, \ell}^{[1]}$;
\For {$j = 1$ to $j_\mathrm{max}$}
\State \hspace{-3.5ex} Compute \ac{sIC} symbols $\tilde{z}_{\theta:m,k}^{[j]}$, $\tilde{z}_{m,\ell}^{t[j]}$ via eq. \eqref{eq:del_t_soft_IC};
\State \hspace{-3.5ex} Compute conditional variances $\sigma_{\theta:m,k}^{2\,[j]}, \sigma_{t:m,\ell}^{2\,[j]}$ via eq. \eqref{eq:del_t_theta_var};
\State \hspace{-3.5ex} Compute extrinsic means $\bar{\theta}_{m,k}^{[j]}$, $\bar{t}_{m,\ell}^{[j]}$ via eq. \eqref{eq:del_t_theta_extr_mean};
\State \hspace{-3.5ex} Compute extrinsic variances $\bar{v}_{\theta:m,k}^{[j]}$, $\bar{v}_{t:m,\ell}^{[j]}$ via eq. \eqref{eq:del_t_theta_extr_var}; \vspace{-0.5ex}
\State \hspace{-3.5ex} Denoise the beliefs $\check{\theta}_{m,k}, \check{t}_{m,\ell}$ via eq. \eqref{eq:del_t_soft_est}; \vphantom{ $\check{x}_{m, k}^{[j]}$}  \vspace{-0.5ex}
\State \hspace{-3.5ex} Denoise the error variances $\check{\psi}_{\theta:m,k} \check{\psi}_{t:m,\ell}$ via eq. \eqref{eq:del_t_est_mse}; \vphantom{ $\check{x}_{m, k}^{[j]}$}  \vspace{-0.5ex}
\State \hspace{-3.5ex} Update the soft-replicas with damping as in eq. \eqref{eq:damped_update}; \vphantom{ $\check{x}_{m, k}^{[j]}$} 
\EndFor
\State Obtain final consensus estimates $\tilde{\theta}_{k}, \tilde{t}_{\ell}$ via eq. \eqref{eq:del_t_final_est};
\State Obtain interference-cancelled system via eq. \eqref{eq:new_lin_syst};
\For {$j = 1$ to $j_\mathrm{max}$}  \vspace{-0.2ex}
\State \hspace{-3.5ex} Compute \ac{sIC} symbols $\tilde{z}{'}_{\!\!\theta:m,k}^{[j]}$ via eq. \eqref{eq:new_lin_SIC};  \vspace{-0.5ex}
\State \hspace{-3.5ex} Compute conditional variances $\sigma_{\theta:m,k}^{2\,[j]}$ via eq. \eqref{eq:new_lin_condvar};  \vspace{-0.5ex}
\State \hspace{-3.5ex} Compute extrinsic means $\bar{\theta}_{m,k}^{[j]}$ via eq. \eqref{eq:del_the_extr_mean};  \vspace{-0.5ex}
\State \hspace{-3.5ex} Compute extrinsic variances $\bar{v}_{\theta:m,k}^{[j]}$ via eq. \eqref{eq:del_the_extr_var};  \vspace{-0.75ex}
\State \hspace{-3.5ex} Denoise the beliefs $\check{\theta}_{m,k}$ via eq. \eqref{eq:del_t_soft_est};  \vspace{-0.5ex} \vphantom{ $\check{x}_{m, k}^{[j]}$}
\State \hspace{-3.5ex} Denoise the error variances $\check{\psi}_{\theta:m,k}$ via eq. \eqref{eq:del_t_est_mse}; \vphantom{ $\check{x}_{m, k}^{[j]}$} \vspace{-0.75ex}
\State \hspace{-3.5ex} Update the soft-replicas with damping as in eq. \eqref{eq:damped_update}; \vphantom{ $\check{x}_{m, k}^{[j]}$} 
\EndFor
\State Obtain refined consensus estimates $\tilde{\theta}_{k}$ via eq. \eqref{eq:del_t_final_est_theta};
\vspace{-2.5ex}
\end{algorithmic} 
\hspace*{\algorithmicindent}
\end{algorithm}

In order to address the aforementioned error behavior of the rotation angle parameters $\boldsymbol{\theta}$, we propose an interference cancellation-based approach to remove the components corresponding to the translation of the sensors, and perform the \ac{GaBP} again only on the rotation angle parameters.
Namely, by using the estimated consensus translation vector $\tilde{\boldsymbol{t}} \triangleq [\tilde{t}_1, \tilde{t}_2, \tilde{t}_3]\trans \!\in \mathbb{R}^{3\times 1}$ obtained at the end of the \ac{GaBP} via equation \eqref{eq:del_t_final_est_t}, the interference-cancelled system is given by
\begin{equation}
\label{eq:new_lin_syst}
\boldsymbol{z}_{n}' \triangleq \boldsymbol{z}_{n} - \boldsymbol{H}_{t} \tilde{\boldsymbol{t}} = \boldsymbol{H}_{\theta} \boldsymbol{\theta} + \boldsymbol{\xi}_{n} \in \mathbb{R}^{M \times 1}.
\end{equation}

The \ac{GaBP} procedure to estimate the rotation angle parameters $\boldsymbol{\theta}$ is identical to the linear \ac{GaBP} of Algorithm \ref{alg:pos_GaBP}, except the factor node equations given by
\begin{equation}
\label{eq:new_lin_SIC}
\tilde{z}{'}_{\!\!\theta:m,k}^{[j]} = z'_{m} - \sum_{i \neq k} h_{\theta:m,i}\hat{\theta}_{m,i}^{[j]} \in \mathbb{R},
\end{equation}
\begin{equation}
\label{eq:new_lin_condvar}
\sigma_{\theta:m,k}^{2\,[j]} = \sum_{i \neq k} \big|h_{\theta:m,i}\big|^2\psi_{\theta:m,i}^{[j]} + N_{0} \in \mathbb{R},
\end{equation}
which is concatenated with the previously described bivariate \ac{GaBP} to describe the complete estimation process of the rigid body transformation parameters $\boldsymbol{\theta}$ and $\boldsymbol{t}$, as summarized by Algorithm \ref{alg:RBL_GaBP}.



\section{Proposed Rigid Body Motion Estimation}
\label{sec:moving_RBL}

\subsection{Velocity Transformation Parameter-based System Model}
\label{sec:reformulation_mov}

In this section, following steps similar to those of the reformulation of the system for stationary \ac{RBL} carried out in the previous section, we reformulate the fundamental system of equation \eqref{eq:linear_sys_mov} to express in terms of the moving \ac{RBL} transformation parameters, \textit{i.e.,} the \ac{3D} angular velocity $\boldsymbol{\omega}\triangleq\left[\omega_{1}, \omega_{2}, \omega_{3}\right]\trans\in\mathbb{R}^{3\times 1}$ and translational velocity vector $\boldsymbol{\dot{t}}\in\mathbb{R}^{3\times 1}$  \cite{ZhaMRBL2021}, in order to enable their estimation.

Utilizing the same small-angle approximation \cite{Diebel2006RigidBodyAttitude} as before, the angular velocity matrix $[\boldsymbol{\omega}]^{\times}$ in equation \eqref{eq:omega_matrix} can be vectorized and decomposed into a linear system directly in terms of the individual velocities \cite{ChenTSP2015}, yielding
\begin{align}
\label{eq:omega_vec}
\mathrm{vec}([\boldsymbol{\omega}]^{\times}) &= \boldsymbol{\Phi} \boldsymbol{\omega} ~\in \mathbb{R}^{9 \times 1} \\
&= \underbrace{
\begin{bmatrix}
0 & 0 & 0 & 0 & 0 & 1 & 0 & -1 & 0 \\
0 & 0 & -1 & 0 & 0 & 0 & 1 & 0 & 0 \\
0 & 1 & 0 & -1 & 0 & 0 & 0 & 0 & 0 \\ 
\end{bmatrix}\trans\!\!\!}_{\triangleq\, \boldsymbol{\Phi} \, \in \, \mathbb{R}^{9 \times 3}} 
\cdot\!
\begin{bmatrix}
\omega_1 \\ \omega_2 \\ \omega_3
\end{bmatrix}\!\!. \nonumber
\end{align}

Then, substituting equation \eqref{eq:omega_vec} into equations \eqref{eq:basic_model_RB_Mov} and \eqref{eq:vel_lin_eq} and rearranging the terms, the following alternate representation of the composite noise is obtained
\begin{align}
\zeta_{m,n} & =\tilde{d}_{m,n} \tilde{\nu}_{m,n} \!-\!\boldsymbol{s}_{n}\trans \boldsymbol{\dot{s}}_{n}\!+\! \boldsymbol{a}_{m}\trans \left( [\boldsymbol{\omega}]^{\times} \boldsymbol{Q} \boldsymbol{c}_{n}+\boldsymbol{\dot{t}}\right) \label{eq:delta_lin_eq_mov} \\
& =\tilde{d}_{m,n} \tilde{\nu}_{m,n} \!-\!\boldsymbol{s}_{n}\trans \boldsymbol{\dot{s}}_{n}\!+\! \left((\boldsymbol{Q} \boldsymbol{c}_{n})\trans \otimes \boldsymbol{a}_{m}\trans\right)\boldsymbol{\Phi}\boldsymbol{\omega}+ \boldsymbol{a}_{m}\trans\boldsymbol{\dot{t}} \in \mathbb{R}, \nonumber
\end{align}
where the matrix product vectorization identity $\mathrm{vec}(\mathbf{X Y Z}) = (\mathbf{Z}\trans \otimes \mathbf{X}) \mathrm{vec}(\mathbf{Y})$ has been used.

In light of the above, similar as before, the fundamental system can be rewritten leveraging the linearization of equation \eqref{eq:delta_lin_eq_mov}, leading to
\begin{subequations}
\label{eq:delta_t_lin_syst_mov}
\begin{equation}
\boldsymbol{u}_{n} \!=\!\! \left[\!\!\begin{array}{c}
\tilde{d}_{1,n} \tilde{\nu}_{1,n} -\boldsymbol{s}_{n}\trans \dot{\boldsymbol{s}}_{n} \\[-1ex]
\vdots \\
\tilde{d}_{M,n} \tilde{\nu}_{M,n} -\boldsymbol{s}_{n}\trans \dot{\boldsymbol{s}}_{n}
\end{array}\!\!\right] = \boldsymbol{B}_{\omega} \!\cdot\! \boldsymbol{\omega} + \boldsymbol{B}_{\dot{t}} \!\cdot\! \dot{\boldsymbol{t}} + \boldsymbol{\zeta}_{n} \in \mathbb{R}^{M \times 1},
\end{equation}
with
\begin{align}
\boldsymbol{B}_{\omega} \!\triangleq\!\! \left[\begin{array}{c}
-\left((\boldsymbol{Q} \boldsymbol{c}_{n})\trans \otimes \boldsymbol{a}_{1}\trans\right)\boldsymbol{\Phi}\\
\vdots   \\[0.5ex]
-\left((\boldsymbol{Q} \boldsymbol{c}_{n})\trans \otimes \boldsymbol{a}_{M}\trans\right)\boldsymbol{\Phi}
\end{array}\right] \!\! \in \!\mathbb{R}^{M \times 3}\!, 
\end{align}
and
\begin{align}
\boldsymbol{B}_{\dot{t}} \!\triangleq\!\! \left[\begin{array}{c}
-\boldsymbol{a}_{1}\trans  \\
\vdots \\
-\boldsymbol{a}_{M}\trans
\end{array}\right]\!\! \in \!\mathbb{R}^{M \times 3}\!,
\end{align}
\end{subequations}
where $\boldsymbol{u}_{n} \in \mathbb{R}^{M \times 1}$ is the effective observed data vector, and $\boldsymbol{B}_{\omega} \in \mathbb{R}^{M \times 3}$ and $\boldsymbol{B}_{\dot{t}} \in \mathbb{R}^{M \times 3}$ are respectively the effective channel matrices for the unknown angular velocity and translational velocity variables, while $\boldsymbol{\zeta}_{n} \in \mathbb{R}^{M \times 1}$ is the vector of composite noise variables from equation \eqref{eq:vel_lin_eq}.

%

\subsection{Linear GaBP for Sensor Velocity Estimation}

From the similarity of the structure of the systems in Sections \ref{sec:reformulation_mov} and Section \ref{sec:pos_est_sol}, it is evident that a linear \ac{GaBP} can be applied to each $n$-th sensor to form a system of corresponding linear equations.
Again, for the sake of notational simplicity, we shall drop subscript $_n$ while stressing that the derivations to follow apply to each $n$-th sensor node.
To be precise, the estimation of the velocity vector $\boldsymbol{\dot{x}}_n$  corresponding to any given $n$-th sensor node in equation \eqref{eq:linear_sys_mov}, here temporarily denoted by $\boldsymbol{\dot{x}} \triangleq [\dot{x}_1, \ldots, \dot{x}_{K+1}]\trans \in \mathbb{R}^{(K+1) \times 1}$, with $k \in \{1,\ldots,K+1\}$ where $K$ is the dimension of the space, is given by the collection of soft replicas for each of its $k$-th element $\dot{x}_{k}$ for each $m$-th observation, which are denoted by $\hat{\dot{x}}_{m,k}^{[j]}$, whose \ac{MSE} is defined as
\begin{equation}
\label{eq:vel_MSE}
\dot{\psi}_{m,k}^{[j]} = \mathbb{E}_{\mathsf{\dot{x}}_{k}}\!\!\left\{\big|\dot{x}_{k}-\hat{\dot{x}}_{m, k}^{[j]}\big|^{2}\right\}=\mathbb{E}_{\mathsf{\dot{x}}_{k}}\!\!\left\{\dot{x}_{k}^{2}\right\}-\hat{\dot{x}}_{m, k}^{2\,[j]} \in \mathbb{R},
\end{equation}
where the superscript $(\,\cdot\,)^{[j]}$ denotes the variable at $j$-th iteration of the \ac{GaBP}.

First, the \ac{sIC} on the $m$-th element of the observed data $y_{m}$ for the estimation of the $k$-th estimated variable is described by
\begin{equation}
\label{eq:vel_soft_IC}
\begin{aligned}
\tilde{\dot{y}}_{m,k}^{[j]} &=\dot{y}_{m}-\sum_{i \neq k} \dot{g}_{m, i} \hat{\dot{x}}_{m, i}^{[j]} \\
&=\dot{g}_{m, k} \dot{x}_{k}+\underbrace{\sum_{i \neq k} \dot{g}_{m, i}\Big(\dot{x}_{i}-\hat{\dot{x}}_{m, i}^{[j]}\Big)+\xi_{m}}_{\triangleq \, \beta_{m,k}^{[j]} \,  \in \, \mathbb{R}} \in \mathbb{R},
\end{aligned}
\end{equation}
where $\tilde{\dot{y}}_{m,k}^{[j]} \in \mathbb{R}$ is the \ac{sIC} symbol corresponding to the $m$-th element of the observed data vector and the $k$-th element of the velocity vector and $\dot{g}_{m,k} \in \mathbb{R}$ is the $(m,k)$-th element of the effective channel matrix $\boldsymbol{\dot{G}}$ in equation \eqref{eq:vel_lin_eq}.

It can be observed that the \ac{GaBP} steps are equivalent to the ones presented in Section \ref{sec:pos_est_sol}.
In what follows, we shall therefore be brief and {\color{black}describe} the steps adjusted to the new corresponding linear system of equations {\color{black}succinctly}. 

%
The conditional \ac{PDF} of $\tilde{\dot{y}}_{\dot{x}:m,k}^{[j]}$ is given  by
\begin{equation}
p_{\tilde{\mathsf{\dot{y}}}_{m,k}^{[j]} \mid \mathsf{\dot{x}}_{k}} \!\Big(\tilde{\dot{y}}_{m,k}^{[j]} \big| \dot{x}_{k}\Big) \propto \mathrm{exp} \! \bigg[-\frac{\big|\tilde{\dot{y}}_{m,k}^{[j]}-\dot{g}_{m, k} \dot{x}_{k}\big|^{2}}{ \sigma_{\dot{x}:m,k}^{2[j]}}\bigg]\!, \label{eq:vel_cond_PDF}
\end{equation}
whose conditional variance is given by
\begin{equation}
\label{eq:vel_var}
\sigma_{m,k}^{2[j]}=\sum_{i \neq k} |\dot{g}_{m, i}|^{2} \dot{\psi}_{m, i}^{[j]} + N_{0} \in \mathbb{R},
\end{equation}
where $N_{0}$ is the power of the composite noise $\xi_n$ described in equation \eqref{eq:delta_lin_eq_mov}.

%
The extrinsic \ac{PDF} of $\dot{x}_{k}$ is written as
\begin{equation}
\label{eq:vel_extr_PDF}
\prod_{i \neq m} p_{\tilde{\mathsf{\dot{y}}}_{i,k}^{[j]} \mid \mathsf{\dot{x}}_{k}} \!\!\left(\tilde{\dot{y}}_{i,k}^{[j]} \mid \dot{x}_{k}\right) \propto \mathrm{exp}
\! \bigg[-\frac{\big|\dot{x}_{k}-\bar{\dot{x}}_{m,k}^{[j]}\big|^{2}}{\bar{v}_{m,k}^{[j]}}\bigg],
\end{equation}
with the extrinsic mean and variance respectively given by
\begin{subequations}
\label{eq:vel_extr}
\begin{equation}
\label{eq:vel_extr_mean}
\bar{\dot{x}}_{m,k}^{[j]} =\bar{v}_{m,k}^{[j]} \cdot \sum_{i \neq m} \frac{\dot{g}_{i, k} \!\cdot\! \tilde{\dot{y}}_{i,k}^{[j]}}{\sigma_{i,k}^{2[j]}}  \in \mathbb{R},
\end{equation}
\begin{equation}
\label{eq:vel_extr_var}
\bar{v}_{m,k}^{[j]} =\bigg(\sum_{i \neq m} \frac{|\dot{g}_{i, k}|^{2}}{\sigma_{i,k}^{2[j]}}\bigg)^{\!\!-1} \!\!\!\!\!\in \mathbb{R}.
\end{equation}
\end{subequations}

%
The denoised mean and variance are given by
\begin{equation}
\label{eq:vel_denoised}
\check{\dot{x}}_{m, k}^{[j]}=\frac{\phi_{\dot{x}} \bar{\dot{x}}_{m, k}^{[j]}}{\phi_{\dot{x}} +\bar{v}_{m,k}^{[j]}} \in \mathbb{R} ~\,\text{and}\,~
\check{\dot{\psi}}_{m,k}^{[j]}=\frac{\phi_{\dot{x}} \bar{v}_{m,k}^{[j]}}{\phi_{\dot{x}}+\bar{v}_{m,k}^{[j]}}\in \mathbb{R}. \!\!\!
\end{equation}

%
The $(j+1)$-th soft-replica is obtained via
\begin{subequations}
\begin{align}
\hat{\dot{x}}_{m, k}^{[j+1]} = \rho \hat{\dot{x}}_{m, k}^{[j]} + (1 - \rho) \check{\dot{x}}_{m, k}^{[j]}, \\[1ex]
\dot{\psi}_{m, k}^{[j+1]} = \rho {\dot{\psi}}_{m, k}^{[j]} + (1 - \rho) \check{\dot{\psi}}_{m, k}^{[j]}.
\end{align}
\label{eq:damped_update_vel}
\end{subequations}
%

$~$
\vspace{-5ex}

Finally, the estimate of the position variable is obtained via
\begin{equation}
\label{eq:vel_final_est}
\tilde{\dot{x}}_{k} = \bigg( \sum_{m = 1}^{M} \frac{|\dot{g}_{m, k}|^2}{\sigma_{m,k}^{2[j_\mathrm{max}]}} \bigg)^{\!\!\!-1} \!\!\!\bigg( \sum_{m = 1}^{M} \frac{\dot{g}_{m, k} \cdot \tilde{\dot{y}}_{m,k}^{[j_\mathrm{max}]}}{ \sigma_{m,k}^{2[j_\mathrm{max}]}} \bigg) \in \mathbb{R}.
\vspace{1ex}
\end{equation}

Algorithm \ref{alg:vel_GaBP} summarizes the proposed method for the sensor velocity estimation without rigid body conformation, based on the range information between the sensors and anchors, ultimately yielding $(K+1)$ elements  in $\boldsymbol{\dot{x}}_n$ composed of $K$ velocity estimates of $\mathbf{\dot{s}}_n$ and the multiplication of sensor position and velocity $\boldsymbol{s}_n\trans\boldsymbol{\dot{s}}_n$.

\begin{algorithm}[t]
\caption{: Linear \ac{GaBP} for Velocity Estimation}
\label{alg:vel_GaBP}
\hspace*{\algorithmicindent}
\begin{algorithmic}[1]
\vspace{-0.9ex}
\Statex \hspace{-4ex} \textbf{Input:} $\boldsymbol{\dot{y}}_n \!~\forall n, \; \boldsymbol{\dot{G}}, \;\phi_{\dot{x}}, \;N_0, \;j_\mathrm{max}, \;\rho$. \vspace{-1.25ex}
\Statex \hspace{-4.4ex} \hrulefill
\Statex \hspace{-4ex}  \textbf{Output:} $\tilde{\dot{x}}_{k} \!~\forall k$ (for all sensor nodes $\forall n$); \vspace{-1.25ex}
\Statex \hspace{-4.4ex} \hrulefill
\Statex \hspace{-3.2ex} \textit{{Perform}} $\forall n, m, k:$ \vspace{0.25ex}
\State Initialize $\hat{\dot{x}}_{m, k}^{[1]}$ and $\dot{\psi}_{m, k}^{[1]}$;
\For {$j = 1$ to $j_\mathrm{max}$}
\State \hspace{-1.5ex} Compute the \ac{sIC} symbol $\tilde{\dot{y}}_{m,k}^{[j]}$ via eq. \eqref{eq:vel_soft_IC};
\State \hspace{-1.5ex} Compute the conditional variance $\sigma_{m,k}^{2[j]}$ via eq. \eqref{eq:vel_var};
\State \hspace{-1.5ex} Compute the extrinsic mean $\bar{\dot{x}}_{m,k}^{[j]}$ via eq. \eqref{eq:vel_extr_mean};
\State \hspace{-1.5ex} Compute the extrinsic variance $\bar{v}_{m,k}^{[j]}$ via eq. \eqref{eq:vel_extr_var};
\State \hspace{-1.5ex} Denoise the beliefs $\check{\dot{x}}_{m, k}^{[j]}, \check{\dot{\psi}}_{m, k}^{[j]}$ via eq. \eqref{eq:vel_denoised};
\State \hspace{-1.5ex} Update the soft-replicas with damping via eq. \eqref{eq:damped_update_vel}; \vphantom{ $\check{\dot{x}}_{m, k}^{[j]}$}
\EndFor
\State Obtain final consensus estimate $\tilde{\dot{x}}_{k}$ via eq. \eqref{eq:vel_final_est}; 
\vspace{-2ex}
\end{algorithmic} 
\hspace*{\algorithmicindent}
\end{algorithm}

Subsequently, the estimated multiplication of sensor position and velocity can be used to construct the second linear equation of equation \eqref{eq:delta_t_lin_syst_mov} incorporating the rigid body conformation in terms of the angular velocity and translational velocity, as will be leveraged in the following subsection to derive the second \ac{GaBP} algorithm to directly estimate the moving rigid body transformation parameters.

\vspace{-1ex}
\subsection{Bivariate GaBP for Velocity Parameter Estimation}

Similar to Algorithm \ref{alg:RBL_GaBP} of Section \ref{sec:del_t_est_sol}, in the case of velocity parameter estimation from equation \eqref{eq:delta_t_lin_syst_mov}, there exist two sets of variables $\omega_k$ with $k \in \{1,\ldots,K\}$ and $\dot{t}_\ell$ with $\ell \in \{1,\ldots,K\}$, such that the \ac{GaBP} rules are elaborated separately.
Again, it can be noticed that the individual steps of the \ac{GaBP} solution are identical to the ones in Section \ref{sec:del_t_est_sol}, applied to the new adjusted system model.
Thus, the new steps will shortly be summarized hereafter.

The \ac{sIC} for the angular and translational velocity variables is given by 
\vspace{-1ex}
\begin{subequations}
\label{eq:omega_t_dot_soft_IC}
\begin{align}\label{eq:omega_IC}
\tilde{u}_{\omega:m,k}^{[j]} &= u_{m} - \sum_{i \neq k} b_{\omega:m,i}\hat{\omega}_{m,i}^{[j]} - \sum_{i = 1}^{K} b_{\dot{t}:m,i}\hat{\dot{t}}_{m,i}^{[j]}, \\[-0.5ex]
&= b_{\omega:m,k}\omega_{k} + \sum_{i = 1}^{K} b_{\dot{t}:m,i}(\dot{t}_{i} - \hat{\dot{t}}_{m,i}^{[j]}) \nonumber \\[-0.5ex] & \hspace{17ex}  + \sum_{i \neq k} b_{\omega:m,i}(\omega_{i} - \hat{\omega}_{m,i}^{[j]}) + \zeta_m,  \nonumber
\end{align}
\vspace{-3ex}
\begin{align}
\label{eq:t_dot_IC}
\tilde{u}_{\dot{t}:m,\ell}^{[j]} &= u_{m} - \sum_{i = 1}^{K} b_{\omega:m,i}\hat{\omega}_{m,i}^{[j]} - \sum_{i \neq \ell}  b_{\dot{t}:m,i}\hat{\dot{t}}_{m,i}^{[j]}, \\[-0.5ex]
&=  b_{\dot{t}:m,\ell}\dot{t}_{\ell} + \sum_{i = 1}^{K}  b_{\omega:m,i}(\omega_{i} - \hat{\omega}_{m,i}^{[j]}) \nonumber \\[-0.5ex] &\hspace{15ex} + \sum_{i \neq \ell}  b_{\dot{t}:m,i}(\dot{t}_{i} - \hat{\dot{t}}_{m,i}^{[j]}) + \zeta_m. \nonumber
\end{align}
\end{subequations}

In turn, the \ac{sIC} symbol conditional \acp{PDF} are given by
\begin{equation}
\label{eq:omega_t_dot_cond_PDF}
\begin{aligned}
p_{\tilde{\mathrm{u}}_{\omega:m,k}^{[j]} \mid \mathrm{\omega}_{k}}(\tilde{u}_{\omega:m,k}^{[j]}|\omega_{k}) &\propto \mathrm{exp}\bigg[ -\frac{|\tilde{u}_{\omega:m,k}^{[j]} - b_{\omega:m,k} \omega_{k}|^2}{\sigma_{\omega:m,k}^{2\,[j]}} \bigg], \\
p_{\tilde{\mathrm{u}}_{\dot{t}:m,\ell}^{[j]} \mid \mathrm{\dot{t}}_{\ell}}(\tilde{u}_{\dot{t}:m,\ell}^{[j]}|\dot{t}_{\ell}) &\propto \mathrm{exp}\bigg[ -\frac{|\tilde{u}_{\dot{t}:m,\ell}^{[j]} - b_{\dot{t}:m,\ell} \dot{t}_{\ell}|^2}{\sigma_{\dot{t}:m,\ell}^{2\,[j]}} \bigg],
\end{aligned}
\end{equation}
with the corresponding conditional variances defined as
\begin{subequations}
\label{eq:omega_t_dot_theta_var}
\begin{align}
\nonumber
\sigma_{\omega:m,k}^{2\,[j]} &= \sum_{i \neq k} \big|b_{\omega:m,i}\big|^2\psi_{\omega:m,i}^{[j]} + \sum_{i = 1}^{K} \big|b_{\dot{t}:m,i}\big|^2\psi_{\dot{t}:m,i}^{[j]} + N_{0},\\[-3ex]\label{eq:omega_theta_var} \\[-1ex]
\nonumber
\sigma_{\dot{t}:m,\ell}^{2\,[j]} &= \sum_{i = 1}^{K} \big|b_{\omega:m,i}\big|^2\psi_{\omega:m,i}^{[j]} + \sum_{i \neq \ell} \big|b_{\dot{t}:m,i}\big|^2\psi_{\dot{t}:m,i}^{[j]}+ N_{0},\\[-3ex]
\label{eq:omega_t_dot_var}
\end{align}
and the corresponding \acp{MSE} $\psi_{\omega:m,k}^{[j]} = \mathbb{E}_{\mathsf{\omega}_k}\!\big[ | \omega_{k} - \hat{\omega}_{m,k}^{[j]} |^2 \big]$ and $\psi_{\dot{t}:m,\ell}^{[j]} = \mathbb{E}_{\mathsf{\dot{t}}_{\ell}}\!\big[ | \dot{t}_{\ell} - \hat{\dot{t}}_{m,\ell}^{[j]} |^2 \big]$.
\end{subequations}

\vspace{1ex}
The extrinsic {PDF} is obtained as
\begin{equation}
\begin{aligned}
\prod_{i \neq m} p_{\tilde{\mathsf{u}}_{\omega:i,k}^{[j]} \mid \mathsf{\omega}_{k}}\left(\tilde{u}_{\omega:i,k}^{[j]} \mid \omega_{k}\right) &\propto \mathrm{exp}\bigg[ -\frac{|\omega_{k} - \bar{\omega}_{m,k}^{[j]}|^2}{\bar{v}_{\omega:m,k}^{[j]}} \bigg], \\[-0.5ex]
\prod_{i \neq m} p_{\tilde{\mathsf{u}}_{\dot{t}:i,\ell}^{[j]} \mid \mathrm{\dot{t}}_{\ell}}\left(\tilde{u}_{\dot{t}:i,\ell}^{[j]} \mid \dot{t}_{\ell}\right) &\propto \mathrm{exp}\bigg[ -\frac{|\dot{t}_{\ell} - \bar{\dot{t}}_{m,\ell}^{[j]}|^2}{\bar{v}_{\dot{t}:m,\ell}^{[j]}} \bigg],
\end{aligned}
\label{eq:omega_t_dot_extr_PDF}
\end{equation}
where the corresponding extrinsic means and variances are defined to be \vspace{-1ex}
\begin{subequations}
\label{eq:omega_t_dot_theta_extr_mean}
\begin{align}
\bar{\omega}_{m,k}^{[j]} &= \bar{v}_{\omega:m,k}^{[j]} \bigg( \sum_{i \neq m} \frac{b_{\omega:i,k} \cdot \tilde{u}_{\omega:i,k}^{[j]}}{ \big(\sigma_{\omega:i,k}^{[j]}\big)^2} \bigg)\in \mathbb{R}, \label{eq:omega_the_extr_mean}\\
\bar{\dot{t}}_{m,\ell}^{[j]} &= \bar{v}_{\dot{t}:m,\ell}^{[j]} \bigg( \sum_{i \neq m} \frac{b_{\dot{t}:i,\ell} \cdot \tilde{u}_{\dot{t}:i,\ell}^{[j]}}{ \big(\sigma_{\dot{t}:i,\ell}^{[j]}\big)^2} \bigg)\in \mathbb{R}, \label{eq:omega_t_dot_extr_mean}
\end{align}
\end{subequations}
\vspace{-2ex}
\begin{subequations}
\label{eq:omega_t_dot_theta_extr_var}
\begin{align}
\bar{v}_{\omega:m,k}^{[j]} &= \bigg( \sum_{i \neq m} \frac{|b_{\omega:i,k}|^2}{\big(\sigma_{\omega:i,k}^{[j]}\big)^2} \bigg)^{\!\!\!-1}  \!\!\!\!\in \mathbb{R}, \label{eq:omega_the_extr_var} \\
\bar{v}_{\dot{t}:m,\ell}^{[j]} &= \bigg( \sum_{i \neq m} \frac{|b_{\dot{t}:i,\ell}|^2}{\sigma_{\dot{t}:m,\ell}^{2\,[j]}} \bigg)^{\!\!\!-1} \!\!\!\!\in \mathbb{R}. \label{eq:omega_t_dot_extr_var}
\end{align}
\end{subequations}

\begin{algorithm}[H]
\caption{: Double {GaBP} for Moving Rigid Body Motion Parameter Estimation}\label{alg:RBL_GaBP_mov}
\hspace*{\algorithmicindent}
\begin{algorithmic}[1]
\vspace{-0.9ex}
\Statex \hspace{-4ex} \textbf{Input:} $\boldsymbol{u}_n \,\boldsymbol{s}_n, \boldsymbol{\dot{s}}_n \!~\forall n,  \boldsymbol{B}_{\omega}, \boldsymbol{B}_{\dot{t}}, \phi_{\omega}, \phi_{\dot{t}}, N_0, j_\mathrm{max}, \rho$. \vspace{-1.25ex}
\Statex \hspace{-4.4ex} \hrulefill
\Statex \hspace{-4ex}  \textbf{Output:} $\tilde{\omega}_{k}$ and $\tilde{\dot{t}}_{\ell} ~\forall k,\ell$ (for all sensor nodes $\forall n$); \vspace{-1.5ex}
\Statex \hspace{-4.4ex} \hrulefill 
\Statex \hspace{-3.2ex} \textit{{Perform}} $\forall n, m, k, \ell:$ \vspace{0.25ex}
\State {\color{black}Initialize} $\hat{\omega}_{m, k}^{[1]}$,  $\hat{\dot{t}}_{m, \ell}^{[1]}$, $\psi_{\omega:m, k}^{[1]}$, $\psi_{\dot{t}:m, \ell}^{[1]}$;
\For {$j = 1$ to $j_\mathrm{max}$}
\State \hspace{-3.5ex} Compute \ac{sIC} symbols $\tilde{u}_{\omega:m,k}^{[j]}$, $\tilde{u}_{\dot{t}:m,\ell}^{[j]}$ via eq. \eqref{eq:omega_t_dot_soft_IC};
%
\State \hspace{-3.5ex} Compute extrinsic means $\bar{\omega}_{m,k}^{[j]}$, $\bar{\dot{t}}_{m,\ell}^{[j]}$ via eq. \eqref{eq:omega_t_dot_theta_extr_mean};
\State \hspace{-3.5ex} Compute extrinsic variances $\bar{v}_{\omega:m,k}^{[j]}$, $\bar{v}_{\dot{t}:m,\ell}^{[j]}$ via eq. \eqref{eq:omega_t_dot_theta_extr_var}; \vspace{-0.5ex}
\State \hspace{-3.5ex} Denoise the beliefs $\check{\omega}_{m,k}, \check{\dot{t}}_{m,\ell}$ via eq. \eqref{eq:omega_t_dot_soft_est}; \vphantom{ $\check{x}_{m, k}^{[j]}$}  \vspace{-0.5ex}
\State \hspace{-3.5ex} Denoise the error variances $\check{\psi}_{\omega:m,k}, \check{\psi}_{\dot{t}:m,\ell}$ via eq. \eqref{eq:omega_t_dot_est_mse}; 
\State \hspace{-3.5ex} Update the soft-replicas with damping as in eq. \eqref{eq:damped_update_mov}; \vphantom{ $\check{x}_{m, k}^{[j]}$} 
\EndFor
\State Obtain final consensus estimates $\tilde{\omega}_{k}, \tilde{\dot{t}}_{\ell}$ via eq. \eqref{eq:omega_t_dot_final_est};
\State Obtain interference-cancelled system via eq. \eqref{eq:new_mov_lin_syst};
\For {$j = 1$ to $j_\mathrm{max}$}  \vspace{-0.2ex}
\State \hspace{-3.5ex} Compute \ac{sIC} symbols $\tilde{u}{'}_{\!\!\omega:m,k}^{[j]}$ via eq. \eqref{eq:new_mov_lin_SIC};  \vspace{-0.5ex}
\State \hspace{-3.5ex} Compute conditional variances $\big(\sigma_{\omega:m,k}^{[j]}\big)^{2}$ via eq. \eqref{eq:new_mov_lin_condvar};  \vspace{-0.5ex}
\State \hspace{-3.5ex} Compute extrinsic means $\bar{\omega}_{m,k}^{[j]}$ via eq. \eqref{eq:omega_the_extr_mean};  \vspace{-0.5ex}
\State \hspace{-3.5ex} Compute extrinsic variances $\bar{v}_{\omega:m,k}^{[j]}$ via eq. \eqref{eq:omega_the_extr_var};  \vspace{-0.75ex}
\State \hspace{-3.5ex} Denoise the beliefs $\check{\omega}_{m,k}$ via eq. \eqref{eq:omega_t_dot_soft_est};  \vspace{-0.5ex} \vphantom{ $\check{x}_{m, k}^{[j]}$}
\State \hspace{-3.5ex} Denoise the error variances $\check{\psi}_{\omega:m,k}$ via eq. \eqref{eq:omega_t_dot_est_mse}; \vphantom{ $\check{x}_{m, k}^{[j]}$} \vspace{-0.75ex}
\State \hspace{-3.5ex} Update the soft-replicas with damping as in eq. \eqref{eq:damped_update_mov}; \vphantom{ $\check{x}_{m, k}^{[j]}$} 
\EndFor
\State Obtain refined consensus estimates $\tilde{\omega}_{k}$ via eq. \eqref{eq:omega_t_dot_final_est_theta};
\vspace{-2.5ex}
\end{algorithmic} 
\hspace*{\algorithmicindent}
\end{algorithm}

Finally, the denoisers with a Gaussian prior are given by 
\begin{subequations}
\begin{equation}
\label{eq:omega_t_dot_soft_est}
\begin{aligned}
\check{\omega}_{m,k} &= \frac{\phi_{\omega} \cdot \bar{\omega}_{m,k}^{[j]}}{\phi_{\omega} + \bar{v}_{\omega:m,k}^{[j]}} \in \mathbb{R}, 
& \check{\dot{t}}_{m,\ell} &= \frac{\phi_{\dot{t}} \cdot \bar{\dot{t}}_{m,\ell}^{[j]}}{\phi_{\dot{t}} + \bar{v}_{\dot{t}:m,\ell}^{[j]}} \in \mathbb{R},\\
\end{aligned}
\end{equation}
\begin{equation}
\label{eq:omega_t_dot_est_mse}
\begin{aligned}
\check{\psi}_{\omega:m,k} &= \frac{\phi_{\omega} \cdot \bar{v}_{\omega:m,k}^{[j]}}{\phi^{\omega} + \bar{v}_{\omega:m,k}^{[j]}} \in \mathbb{R},
& \check{\psi}_{\dot{t}:m,\ell} &= \frac{\phi_{\dot{t}} \cdot \bar{v}_{\dot{t}:m,\ell}^{[j]}}{\phi_{\dot{t}} + \bar{v}_{\dot{t}:m,\ell}^{[j]}} \in \mathbb{R},
\end{aligned}
\end{equation}
\end{subequations}

\vspace{1ex}
\noindent where $\phi_{\omega}$ and $\phi_{\dot{t}}$ are the element-wise variances of $\boldsymbol{\omega}$ and $\boldsymbol{\dot{t}}$.

The $(j+1)$-th soft-replica is obtained via a damped update, described by
\begin{subequations}
\begin{align}
\hat{x}_{m, k}^{[j+1]} = \rho \hat{x}_{m, k}^{[j]} + (1 - \rho) \check{x}_{m, k}^{[j]}, \\[1ex]
\psi_{m, k}^{[j+1]} = \rho {\psi}_{m, k}^{[j]} + (1 - \rho) \check{\psi}_{m, k}^{[j]}.
\end{align}
\label{eq:damped_update_mov}
\end{subequations}

Finally, after $j_\mathrm{max}$ iterations the consensus estimates are obtained from 
\begin{subequations}
\label{eq:omega_t_dot_final_est}
\begin{align}
\tilde{\omega}_{k} &= \bigg( \sum_{m = 1}^{M} \frac{|b_{\omega:m,k}|^2}{\big(\sigma_{\omega:m,k}^{[j_\mathrm{max}]}\big)^2} \bigg)^{\!\!\!-1} \! \! \bigg( \sum_{m = 1}^{M} \frac{b_{\omega:m,k} \cdot \tilde{u}_{\omega:m,k}^{[j_\mathrm{max}]}}{ \big(\sigma_{\omega:m,k}^{[j_\mathrm{max}]}\big)^2} \bigg) \in \mathbb{R}, \nonumber \\[-2ex] \label{eq:omega_t_dot_final_est_theta} \\
\tilde{\dot{t}}_{\ell} &= \bigg( \sum_{m = 1}^{M} \frac{|b_{\dot{t}:m,\ell}|^2}{\big(\sigma_{\dot{t}:m,\ell}^{[j_\mathrm{max}]}\big)^2} \bigg)^{\!\!\!-1} \!\bigg( \sum_{m = 1}^{M} \frac{b_{\dot{t}:m,\ell} \cdot \tilde{u}_{\dot{t}:m,\ell}^{[j_\mathrm{max}]}}{\big(\sigma_{\dot{t}m,\ell}^{[j_\mathrm{max}]}\big)^2} \,\bigg) \in \mathbb{R}. \nonumber \\[-2ex] \label{eq:omega_t_dot_final_est_t}
\end{align}
\vspace{-3ex}
\end{subequations}

Using the estimated consensus translational velocity vector $\tilde{\boldsymbol{\dot{t}}} \triangleq [\tilde{\dot{t}}_1, \tilde{\dot{t}}_2, \tilde{\dot{t}}_3]\trans \!\in \mathbb{R}^{3\times 1}$ obtained at the end of the {GaBP} via equation \eqref{eq:del_t_final_est_t}, the interference-cancelled system is given by \vspace{-0.5ex}
\begin{equation}
\label{eq:new_mov_lin_syst}
\boldsymbol{u}_{n}' \triangleq \boldsymbol{u}_{n} - \boldsymbol{B}_{\dot{t}} \tilde{\boldsymbol{\dot{t}}} = \boldsymbol{B}_{\omega} \boldsymbol{\omega} + \boldsymbol{\zeta}_{n} \in \mathbb{R}^{M \times 1}. \vspace{-1ex}
\end{equation}

The {GaBP} procedure to estimate the angular velocity parameters $\boldsymbol{\omega}$ is identical to the linear {GaBP} in Algorithm \ref{alg:vel_GaBP}, where the post-\ac{IC} factor node equations are given by 
\begin{equation}
\label{eq:new_mov_lin_SIC}
\tilde{u}{'}_{\!\!\omega:m,k}^{[j]} = u'_{m} - \sum_{i \neq k} b_{\omega:m,i}\hat{\omega}_{m,i}^{[j]} \in \mathbb{R},
\vspace{-1ex}
\end{equation}
\begin{equation}
\label{eq:new_mov_lin_condvar}
\sigma_{\omega:m,k}^{2\,[j]} = \sum_{i \neq k} \big|b_{\omega:m,i}\big|^2\psi_{\omega:m,i}^{[j]} + N_{0} \in \mathbb{R},
\end{equation}
which is then again concatenated with the previous bivariate {GaBP} to describe the complete estimation process for the rigid body transformation parameters $\boldsymbol{\omega}$ and $\boldsymbol{\dot{t}}$, as summarized by Algorithm \ref{alg:RBL_GaBP_mov}.


\section{Performance Analysis}
\label{sec:performance_analysis}

\subsection{RBL Parameter Estimation Performance Against the SotA}

In this section, we present simulation results to demonstrate the effectiveness of the proposed multi-stage \ac{GaBP}-based approach for \ac{RBL}.
Specifically, we compare the estimation performance of the two stages of the proposed approaches: \textit{a)} the sensor position estimation via linear \ac{GaBP} in Algorithm \ref{alg:pos_GaBP}, and the \ac{RBL} transformation parameter estimation via bivariate \ac{GaBP} and interference cancellation-based refinement \ac{GaBP} concatenated in Algorithm \ref{alg:RBL_GaBP} for a stationary rigid body, and \textit{b)} the sensor velocity estimation via Algorithm \ref{alg:vel_GaBP}, and the \ac{RBL} velocity parameter estimation via Algorithm \ref{alg:RBL_GaBP_mov} for a moving rigid body.
{\color{black}The performance is evaluated in comparison with two relevant \ac{SotA} \ac{RBL} methods. The first method, presented in \cite{ChenTSP2015}, involves a three-stage estimation procedure. Initially, sensor positions are estimated using a \ac{ToA}-based approach as described in \cite{MaICASSP2011}. Subsequently, sensor velocities are jointly estimated from \ac{TDOA} and \ac{FDOA} measurements following the method in \cite{HoTSP2004}. Finally, the \ac{RBL} parameters are estimated using the approach outlined in \cite{ChenTSP2015}.

The second \ac{SotA} method considered is that proposed in \cite{Jiang2019}, which is a high complexity and high accuracy technique. This method estimates the \ac{RBL} parameters using semidefinite relaxation (\ac{SDR}) applied to \ac{ToA} and Doppler measurements.}

The simulation setup is the scenario illustrated in Figure \ref{fig:sys_mod_plot}, where the rigid body is composed of $N = 8$ sensors positioned at the vertices of a unit cube at the origin, with sensor positions described{\color{black}, in meters,} by the conformation matrix given by
\begin{equation*}\label{eq:C_mat}
\boldsymbol{C} \!=\!\!
\resizebox{0.39 \textwidth}{!}{$\left[\begin{array}{lllllllc}
-0.5 &\! \phantom{-}0.5 &\! \phantom{-}0.5 &\! -0.5 &\! -0.5 &\! \phantom{-}0.5 &\! -0.5 &\! \phantom{-}0.5 \\
-0.5 &\!-0.5 &\! \phantom{-}0.5 &\!\phantom{-}0.5 &\! -0.5 &\! -0.5 & \!\phantom{-}0.5 & \!\phantom{-}0.5 \\
-0.5 &\! -0.5 &\! -0.5 &\! -0.5 & \!\phantom{-}0.5 &\! \phantom{-}0.5 &\! \phantom{-}0.5 &\! \phantom{-}0.5
\end{array}\right]
$} \!\!\in\! \mathbb{R}^{3 \times 8}\!,
\end{equation*}
and the $M = 8$ anchors are positioned at the vertices of a larger cube (\textit{i.e.,} room), where the anchor conformation matrix $\boldsymbol{A} \in \mathbb{R}^{3\times 8}$ is given by
\begin{equation*}\label{eq:A_mat}
\boldsymbol{A} \!=\!\!
\resizebox{0.38 \textwidth}{!}{$
\left[\begin{array}{lllllllc}
-10 &\! \phantom{-}10 &\! \phantom{-}10 &\! -10 & \!-10 & \!\phantom{-}10 &\! -10 &\! \phantom{-}10 \\
-10 & \!-10 & \!\phantom{-}10 &\!\phantom{-}10 &\! -10 & \!-10 & \!\phantom{-}10 & \!\phantom{-}10 \\
-10 & \!-10 &\! -10 &\! -10 & \!\phantom{-}10 &\! \phantom{-}10 &\! \phantom{-}10 & \!\phantom{-}10
\end{array}\right]
$}  \!\!\in\! \mathbb{R}^{3 \times 8}\!.
\end{equation*}

The \ac{RBL} rotation angles $\theta_x, \theta_y, \theta_z$ follow a zero-mean Gaussian distribution of variance $\phi_{\theta} = 10 {\color{black}\text{deg}^2}$, and the \ac{RBL} translation vector elements also follow a zero-mean Gaussian distribution of variance $\phi_{t} = 5 {\color{black}\text{m}^2}$.
The \ac{RBL} parameters in the moving scenario follow a zero-mean Gaussian distribution of variance $\phi_{\omega} = 10{\color{black}(\text{deg}/\text{s})^2}$ for the angular velocity and a variance of $\phi_{\dot{t}} = 5 {\color{black}(\text{m}/\text{s})^2}$ for the translational velocity.

\begin{figure}[t]
\centering
{{\includegraphics[width=1\columnwidth]{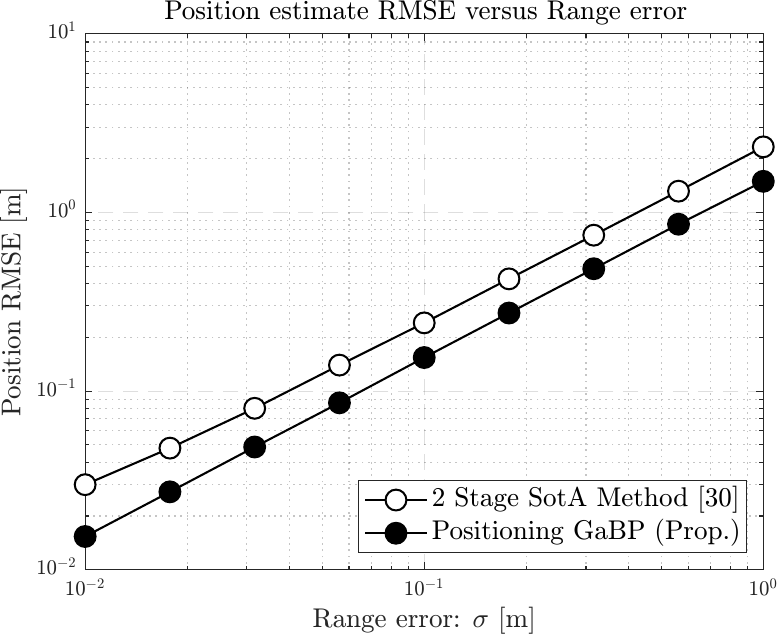}}}    
\vspace{-2ex}
\caption{\Ac{RMSE} of the proposed linear \ac{GaBP} algorithm for sensor position estimation (Alg. \ref{alg:pos_GaBP}) and the \ac{SotA} method of \cite{MaICASSP2011}, for various range errors $\sigma$.}
\label{fig:pos_rmse_plot}
\end{figure}

The performance is assessed using the \ac{RMSE} defined as
\begin{equation}
\mathrm{RMSE} = \sqrt{\frac{1}{E}\sum_{i=1}^{E}\|\hat{\boldsymbol{x}}^{[i]}-{\boldsymbol{x}}\|_2^2},
\end{equation}
where $\hat{\boldsymbol{x}}^{[i]}$ is the \ac{RBL} parameter vector (position, angle, or translation) estimated during the $i$-th Monte-Carlo simulation, $\boldsymbol{x}$ is the true \ac{RBL} parameter vector, and $E = 10^4$ is the total number of independent Monte-Carlo experiments used for the analysis, and is evaluated for different noise standard deviations\footnote{\color{black}In the moving scenario the range and Doppler noise is modeled coupled, such that a varying noise level effects both, the range and Doppler measurements. It is important to note that for certain scenarios a sensitivity analysis can become of interest, where the effect of the different types of noise are investigated individually, which is, however, not applicable to the proposed scenario.} $\sigma$ of equation \eqref{eq:range_model} and \eqref{eq:doppler_model}, defined as {\color{black}$\sigma \!\triangleq\! \sigma_w \!=\!0.1\sigma_\epsilon$.
For the \ac{GaBP} algorithms, the damping factor is set to $\rho = 0.5$, and the maximum number of iterations is set to $j_\mathrm{max} = 30$, which has empirically shown sufficient convergence.}

\begin{figure}[t]
\centering
{{\includegraphics[width=1\columnwidth]{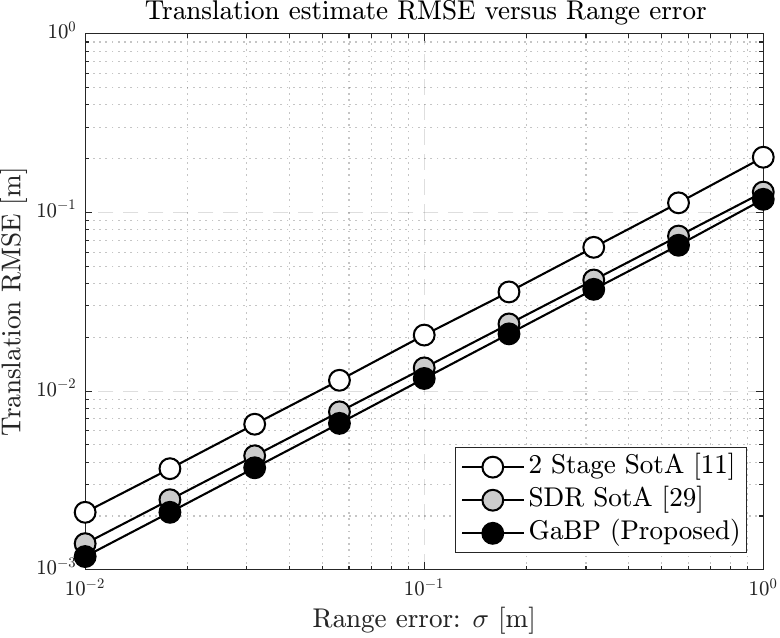}}}
\caption{\Ac{RMSE} of the proposed bivariate \ac{GaBP} algorithm for rigid body translation estimation (Alg. \ref{alg:RBL_GaBP}) and the \ac{SotA} methods of \cite{ChenTSP2015} {\color{black}and \cite{Jiang2019}}, for various range errors $\sigma$.}
\label{fig:tran_rmse_plot}

\vspace{3ex}
{{\includegraphics[width=1\columnwidth]{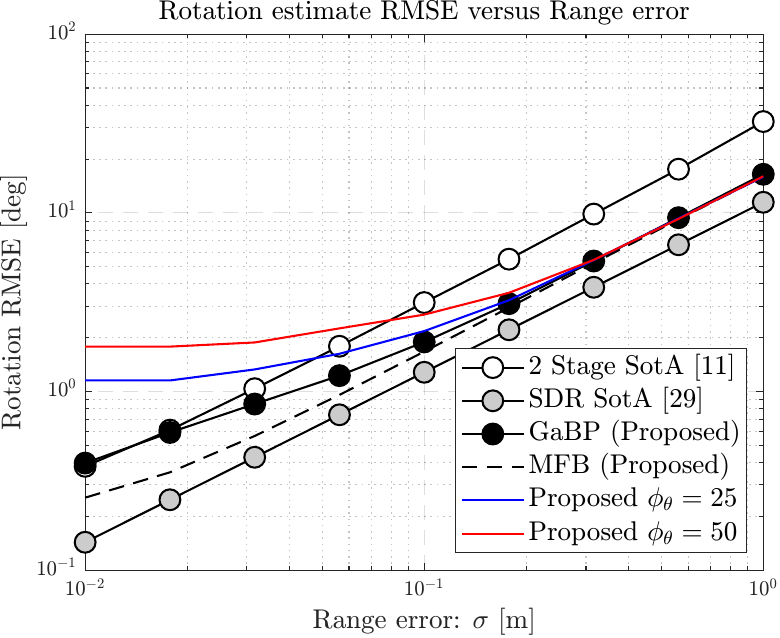}}}
\vspace{-2ex}
\caption{\Ac{RMSE} of the proposed bivariate \ac{GaBP} algorithm for rigid body rotation angle estimation (Alg. \ref{alg:RBL_GaBP}) and the \ac{SotA} methods of \cite{ChenTSP2015} {\color{black}and \cite{Jiang2019}}, for various range errors $\sigma$.}
\label{fig:rot_rmse_plot}
\vspace{-2ex}
\end{figure}

Firstly, for the stationary scenario, Figure \ref{fig:pos_rmse_plot} illustrates the sensor position estimation performance of the proposed Algorithm \ref{alg:pos_GaBP} and the \ac{SotA} closed-form solution of \cite{MaICASSP2011} for different noise standard deviations, also referred to as range error in meters.
It can be observed that the proposed linear \ac{GaBP} method outperforms the \ac{SotA} approach for all range error regimes, which suggests that already the proposed preliminary positioning method can be used as an initializer for other \ac{SotA} methods to perform \ac{RBL} parameter extraction.
In addition, Figures \ref{fig:tran_rmse_plot} and \ref{fig:rot_rmse_plot} illustrate the estimation performance of the rigid body translation vector and rotation angles respectively.

As mentioned in Section \ref{sec:proposed_stationary}, the translation parameter can be effectively estimated via the proposed bivariate \ac{GaBP} of Algorithm \ref{alg:RBL_GaBP} in light of the channel power difference effect, which is highlighted by the result of Figure \ref{fig:tran_rmse_plot}{\color{black}, even outperforming the high accuracy \ac{SDR}-method of \cite{Jiang2019}}.
%

Finally, Figure \ref{fig:rot_rmse_plot} illustrates the estimation performance of the rotation angles of the rigid body transformation.
It can be seen that the estimation performance of the proposed concatenated \ac{GaBP} also exhibits superiority to the \ac{SotA} method of \cite{ChenTSP2015}, similarly to the behavior of the positioning and translation estimation performance{\color{black}, but is slightly inferior to the \ac{SDR}-method \cite{Jiang2019}.}
Due to the aforementioned channel power scaling effect which causes the noise power to become more prominent in the estimation of the variables for small range error regimes, the performance of the proposed method can be seen to approach the performance of the {\color{black}2 Stage} \ac{SotA} in Figure \ref{fig:rot_rmse_plot}.
{\color{black}This is due to error floors} usually seen in belief propagation algorithms at smaller error regions, which can be further optimized via adaptive damping \cite{Vila_2015}.

Additionally, the ideal behavior of the proposed algorithm is illustrated via the \ac{MFB} of the \ac{GaBP} algorithm, {\color{black}where the algorithm soft-replicas are initialized with the true values \cite{Ranasinghe2025JCDE, He2023HybridMP} to illustrate the optimal performance bound,} which shows the expected superiority over all noise ranges, also illustrating that for high noise, the proposed method indeed reaches the optimal performance.
\vspace{-2ex}
\begin{figure}[H]
\centering
{{\includegraphics[width=0.97\columnwidth]{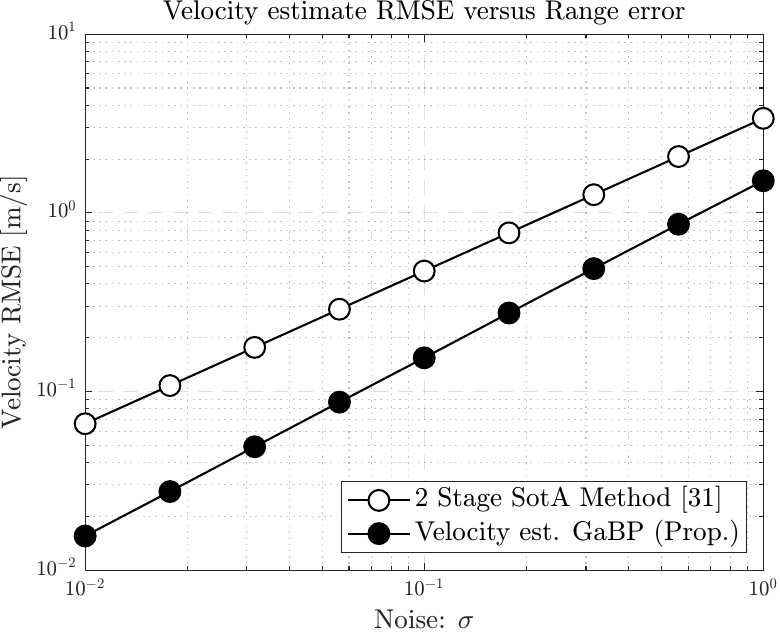}}}
\caption{\Ac{RMSE} of the proposed linear \ac{GaBP} algorithm for sensor velocity estimation (Alg. \ref{alg:vel_GaBP}) and the \ac{SotA} method of \cite{HoTSP2004}, for various noise levels $\sigma$.}
\label{fig:vel_rmse_plot}
\vspace{1ex}
{{\includegraphics[width=0.97\columnwidth]{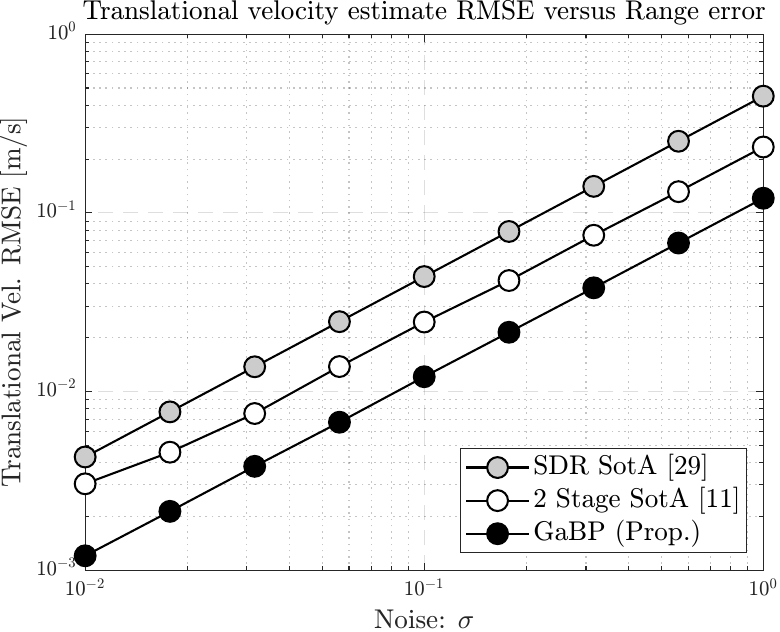}}}
\caption{\Ac{RMSE} of the proposed bivariate \ac{GaBP} algorithm for rigid body translational velocity estimation (Alg. \ref{alg:RBL_GaBP_mov}) and the \ac{SotA} methods of \cite{ChenTSP2015} {\color{black}and \cite{Jiang2019}}, for various noise levels $\sigma$.}
\label{fig:tran_vel_rmse_plot}
\end{figure}

\begin{figure}[H]
\centering
{{\includegraphics[width=0.97\columnwidth]{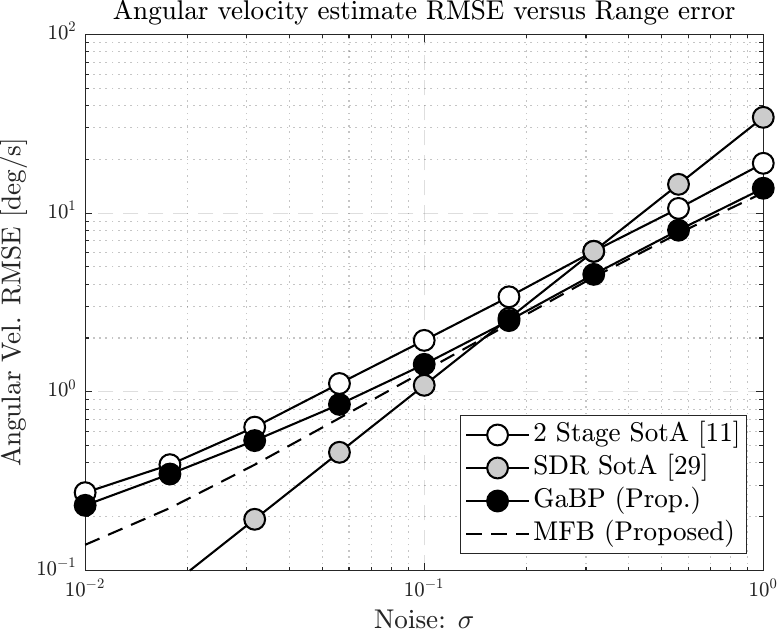}}}   
\caption{\Ac{RMSE} of the proposed bivariate \ac{GaBP} algorithm for rigid body angular velocity estimation (Alg. \ref{alg:RBL_GaBP_mov}) and the \ac{SotA} methods of \cite{ChenTSP2015} {\color{black}and \cite{Jiang2019}}, for various noise levels $\sigma$.}
\label{fig:omega_rmse_plot}
\end{figure}

{\color{black}
Furthermore, it should be noted that the use of the small angle approximation, although valid for the scenario considered in this work, may lead to singularities when the rotation angles become large\footnotemark.
A notable example is the gimbal lock phenomenon, in which two rotation axes align, causing a loss of one degree of freedom in the parameterization.

Even before such issues occur, the Jacobian matrix may become ill conditioned, resulting in numerical instability during the estimation process. While such issues do not manifest in the current simulation setup, scenarios involving high dynamics or large rotation angles may necessitate the adoption of alternative estimation techniques. Approaches based on quaternions \cite{Markley2000} or Lie groups \cite{Barfoot2017} have been proposed to address these challenges, offering robust estimation without susceptibility to singularities.

Finally, it is observed that, even in cases where the rotation angles are large and the small angle approximation is no longer valid, the estimation performance remains comparable to that of smaller angle scenarios under conditions of high range error. However, in low range error regimes, the performance converges to an error floor due to the limitations imposed by the small angle approximation.}


\footnotetext{\color{black}This problem can be mitigated by replacing the linear approximation with a quadratic approach, combined with a strategy whereby the variable squared is treated as another variable correlated to that latter \cite{Rou_2025}. To this end, however, a non-trivial redesign of the GaBP algorithm is required, which is out of the scope of this article and will be pursued in a follow up work.}

\begin{table*}[t]
\centering
\caption{Complexity and Runtime}
\vspace{-1ex}
\begin{tabular}[H]{|l|c|c|}
\hline
\bf Method&\bf Complexity &\bf Runtime\\[0.5ex]
\hline\hline
Prop. Stationary Linear \ac{GaBP} (Alg. 1)&$O(NMK)$& $0.15$ms\\[0.5ex]
\hline
Prop. Moving Linear \ac{GaBP} (Alg. 3)&$O(NMK)$& $0.14$ms\\[0.5ex]
\hline
\ac{SotA} Position/Velocity Est. \cite{HoTSP2004} & $O(MNK^2)$ &$1.6$ms\\[0.5ex]
\hline
\hline
Prop. Stationary Double \ac{GaBP} (Alg. 2)&$O(NMK^2)$& $1.1$ms\\[0.5ex]
\hline
{\color{black}2-Stage} \ac{SotA} Stationary Parameter Est. \cite{ChenTSP2015} & $O(K^3+MN)$ &$0.24$ms\\[0.5ex]
\hline
{\color{black}\ac{SDR}-based \ac{SotA} Stationary Parameter Est. \cite{Jiang2019}} & {\color{black}$O\left((K+1)\left(MN(K+1)^4 + M^2N^2(K+1)^2 + (K+1)^8\right)\right)$
} &{\color{black}$4.6$ms}\\[0.5ex]
\hline\hline
Prop. Moving Double \ac{GaBP} (Alg. 4)&$O(NMK^2)$& $1$ms\\[0.5ex]
\hline
{\color{black}2-Stage} \ac{SotA} Moving Parameter Est. \cite{ChenTSP2015} & $O(K^3+MN)$ &$0.39$ms\\[0.5ex]
\hline
{\color{black}\ac{SDR}-based \ac{SotA} Moving Parameter Est. \cite{Jiang2019}} & {\color{black}$O\left(8\sqrt{2}(K+1)\left(MN(K+1)^4 + M^2N^2(K+1)^2 + 4(K+1)^8\right)\right)$
} &{\color{black}$12.2$ms}\\[0.5ex]
\hline
\end{tabular}
\label{table:1}
\end{table*}

Next, the second set of results evaluates the performance of the proposed rigid body motion estimation algorithms within the moving rigid body scenario.
First, Figure \ref{fig:vel_rmse_plot} illustrates the \ac{RMSE} of the sensor velocity estimates of Algorithm \ref{alg:vel_GaBP} compared to the \ac{SotA} solution of \cite{HoTSP2004}.
It can be observed that the proposed method outperforms the \ac{SotA} in all velocity errors with a large performance gain.
{\color{black}In turn,} Figures \ref{fig:tran_vel_rmse_plot} and \ref{fig:omega_rmse_plot} show the performance of the estimate velocity parameters, $i.e.$, the {\color{black}translational} velocity and the {\color{black}angular} velocity.

\begin{figure}[H]
\centering
{{\includegraphics[width=1\columnwidth]{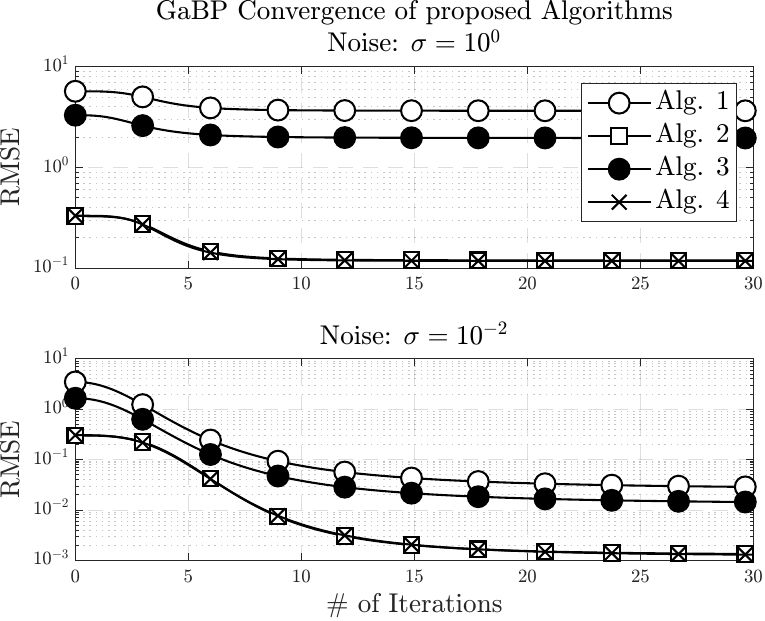}}}
\vspace{-2ex}
\caption{Convergence of the proposed algorithms for varying noise levels of $\sigma=10^{0}$ and $\sigma=10^{-2}$.}
\label{fig:convergence}
\vspace{-1.5ex}
\end{figure}

In Figure \ref{fig:tran_vel_rmse_plot}, the translational velocity estimate of the proposed method in Algorithm \ref{alg:RBL_GaBP_mov} is compared to the \ac{SotA} methods of \cite{ChenTSP2015} {\color{black}and \cite{Jiang2019}}.
It can be observed that as previously, the proposed method outperforms the \ac{SotA} for all range errors.
In turn, Figure \ref{fig:omega_rmse_plot} illustrates the performance of the angular velocity estimation, where similar to the stationary rotation estimation results, the proposed method outperforms the 2 Stage \ac{SotA} of \cite{ChenTSP2015} for all range errors, with a larger gain reaching the \ac{MFB} for high noise and a smaller gain for low noise{\color{black}, while the \ac{SDR}-based method \cite{Jiang2019} outperforms our proposed method for low noise levels.}

\vspace{-1.5ex}
\subsection{Convergence Behavior and Complexity Analysis}

Here, the performance of the proposed \ac{RBL} parameter estimation algorithms are further assessed in terms of the computational complexity and convergence behavior, which is thoroughly investigated in \cite{Su_2014,Su_2015,Li_2019}, and compared against the \ac{SotA} methods.
First, Figure \ref{fig:convergence} illustrates the convergence {\color{black}behavior} of the four proposed algorithms, where it can be observed that all four algorithms converge very efficiently, {\color{black}with the bivariate Algorithms {\color{black}\ref{alg:RBL_GaBP}} and {\color{black}\ref{alg:RBL_GaBP_mov}} converging slightly steeper, and to a higher estimation accuracy than the two linear algorithms.} 
It should also be noted that since the \ac{SotA} algorithms are not iterative, a convergence analysis comparison is not relevant.

Finally, the computational complexities of the proposed algorithms have been outlined in Table \ref{table:1} in terms of the complexity order on the system size parameters using the well-known Big-O notation, as well as a convenient measure of practical runtimes in seconds simulated via an ordinary computer.
{\color{black}It can be observed that the proposed \ac{GaBP} algorithms provide an accurate estimation of the \ac{RBL} parameters with a low computational complexity, comparable to the \ac{SotA} methods of \cite{ChenTSP2015} and \cite{HoTSP2004}, while being far superior to the \ac{SotA} method of \cite{Jiang2019}.}

\section{Conclusion}
\label{sec:conclusion}

We presented a novel and efficient framework for solving the \ac{6D} \ac{RBL} problem, for under stationary and moving \textit{i.e.,} estimation of \ac{3D} location and \ac{3D} rotation, as well as corresponding translation and angular velocities, via a series of tailored \ac{GaBP} message passing estimators.
First, a linear \ac{GaBP} for the sensor position estimation based on the obtained range estimates from the wireless sensors is derived, from which the \ac{RBL} system is reformulated via the small-angle approximation to enable the construction of a second bivariate \ac{GaBP} which is capable of directly estimating the rotation angles and the translation distances. 
Next, Doppler measurements were obtained, designing a system of linear equations that can be used to estimate the sensor velocity via a linear \ac{GaBP}, followed by the estimation of angular and translational velocity by the construction of another bivariate \ac{GaBP}.
The proposed \ac{RBL} method is shown to outperform the \ac{SotA} method in all the position, rotation, translation, sensor velocity, angular velocity and translational velocity estimation performance, in addition to the complexity advantage highlighted by the analysis.
%



{
\color{black}
\begin{IEEEbiography}[{\includegraphics[width=1in,height=1.25in,clip,keepaspectratio]{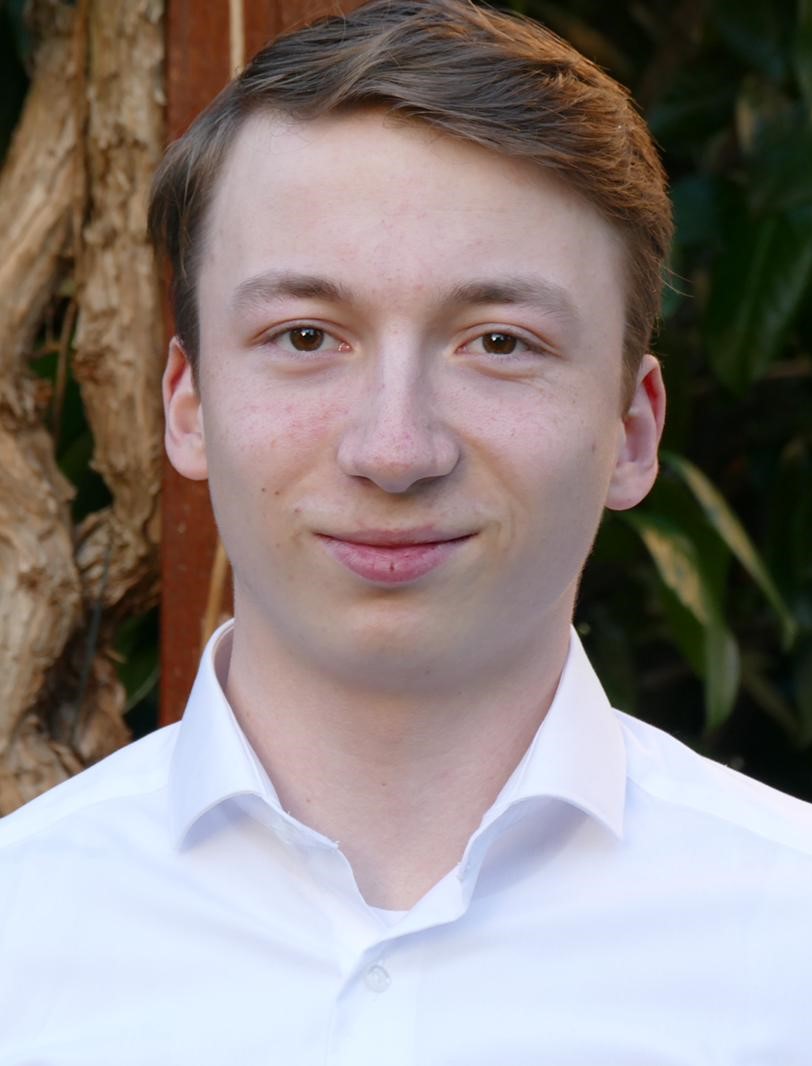}}]{Niclas F{\"u}hrling}
(Graduate Student Member, IEEE) received the B.Sc. degree in electrical and computer engineering from Jacobs University Bremen, Bremen, Germany in 2022 and the M.Sc. degree in electrical engineering with the University of Bremen in 2024, with a focus on communication and information technology. He is currently pursuing the Ph.D. degree in Electrical Engineering with Constructor University Bremen, Germany, while working on a research project  focusing on 6G connectivity. His current research interests are wireless communications, signal processing and wireless localization.
\end{IEEEbiography}
\begin{IEEEbiography}[{\includegraphics[width=1in,height=1.25in,clip,keepaspectratio]{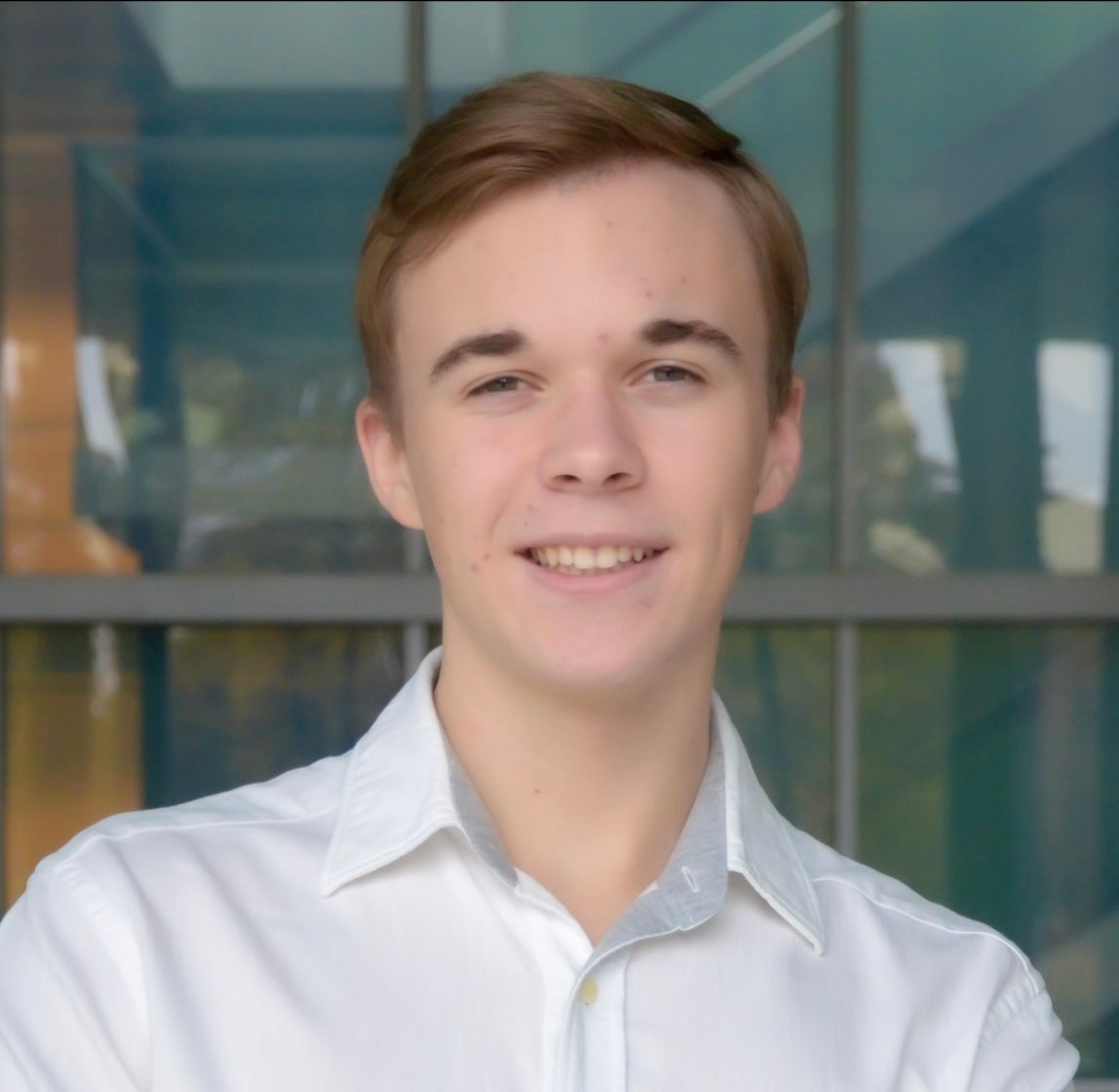}}]{Volodymyr Vizitiv} (Graduate Student Member, IEEE) obtained the B.Sc. degree in Electrical and Computer Engineering at Constructor University Bremen, Germany, in 2024. He is currently pursuing a Ph.D. in Electrical and Computer Engineering at Rice University, Houston, Texas, USA, with a focus on THz wireless communications. His ongoing research interests include metasurface-based wavefront engineering and machine learning–driven inverse design for next-generation THz wireless systems.
\end{IEEEbiography}
\begin{IEEEbiography}[{\includegraphics[width=1in,height=1.25in,clip,keepaspectratio]{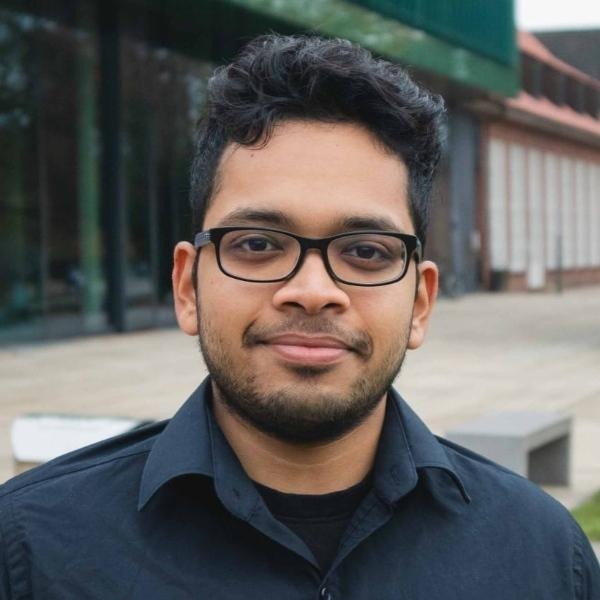}}]{Kuranage Roche Rayan Ranasinghe} (Graduate Student Member, IEEE) received the B.Sc. degree in electrical and computer engineering and the B.Sc. degree in robotics and intelligent systems (with a minor in physics) from Constructor University (formerly Jacobs University Bremen) in 2023 and 2024, respectively, where he is currently pursuing the Ph.D. degree in electrical engineering. He was a visiting researcher at ETH Zürich in Switzerland and at KTH Royal Institute of Technology in Sweden in 2025. Within the fields of wireless communications and signal processing, his research interests encompass integrated sensing, communications and computing (ISCC), compressed sensing, Bayesian inference, next-generation metasurface technologies, and optimization theory. He received the Best Paper Award at the International Conference on Computing, Networking and Communications (ICNC) in 2025 and he also contributes to the IEEE Standards Association P3383 Working Group on Performance Metrics for Integrated Sensing and Communication.
\end{IEEEbiography}
\begin{IEEEbiography}[{\includegraphics[width=1in,height=1.25in,clip,keepaspectratio]{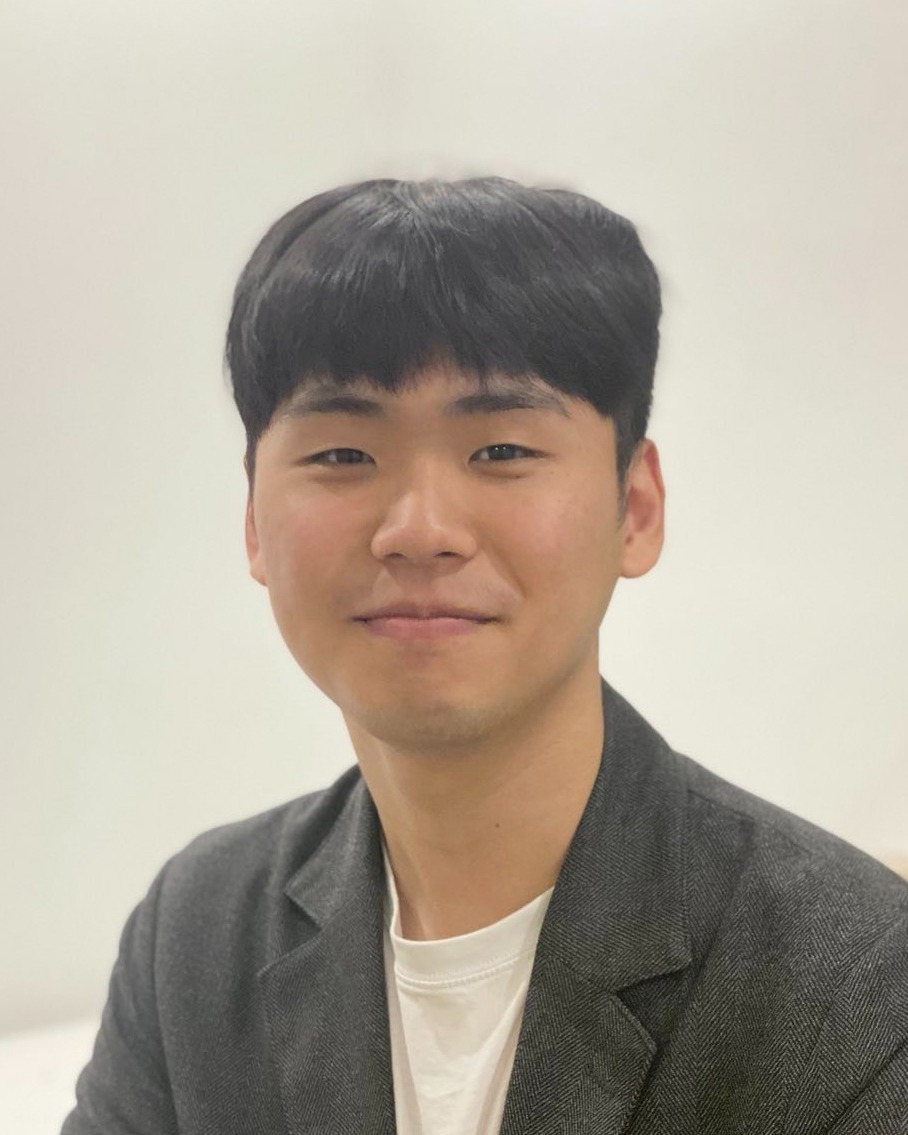}}]{Hyeon Seok Rou} (Member, IEEE) received his Ph.D. in electrical engineering from Constructor University Bremen, Germany, in 2024, and his B.Sc. in electrical and computer engineering from the same institution (formerly Jacobs University Bremen) in 2021. Since 2025, he has been a postdoctoral research fellow and lecturer at the School of Computer Science and Engineering, Constructor University Bremen, Germany. His research interests include next-generation (B5G,6G) wireless systems, integrated sensing and communications (ISAC), affine frequency division multiplexing (AFDM), signal processing in doubly-dispersive channels, high-mobility communications, multi-dimensional modulation, over-the-air computing (AirComp), V2X technologies, and quantum-accelerated optimisation. He has served as the chapter representative of The Korean Scientists and Engineers Association in the FRG (VeKNI e.V.) since 2021, and is a delegate to the Young Professionals Forum (YPF) at the 3rd World Congress of Korean Scientists and Engineers in South Korea, 2025. He was a visiting researcher at the Intelligent Communications Laboratory (ICL!) at KAIST in 2023, and received the Korea Institute of Science and Technology Europe Research Scholarship Award in 2022. He also contributes to the IEEE Standards Association P3383 Working Group on Performance Metrics for Integrated Sensing and Communication.
\end{IEEEbiography}
\begin{IEEEbiography}[{\includegraphics[width=1in,height=1.25in,clip,keepaspectratio]{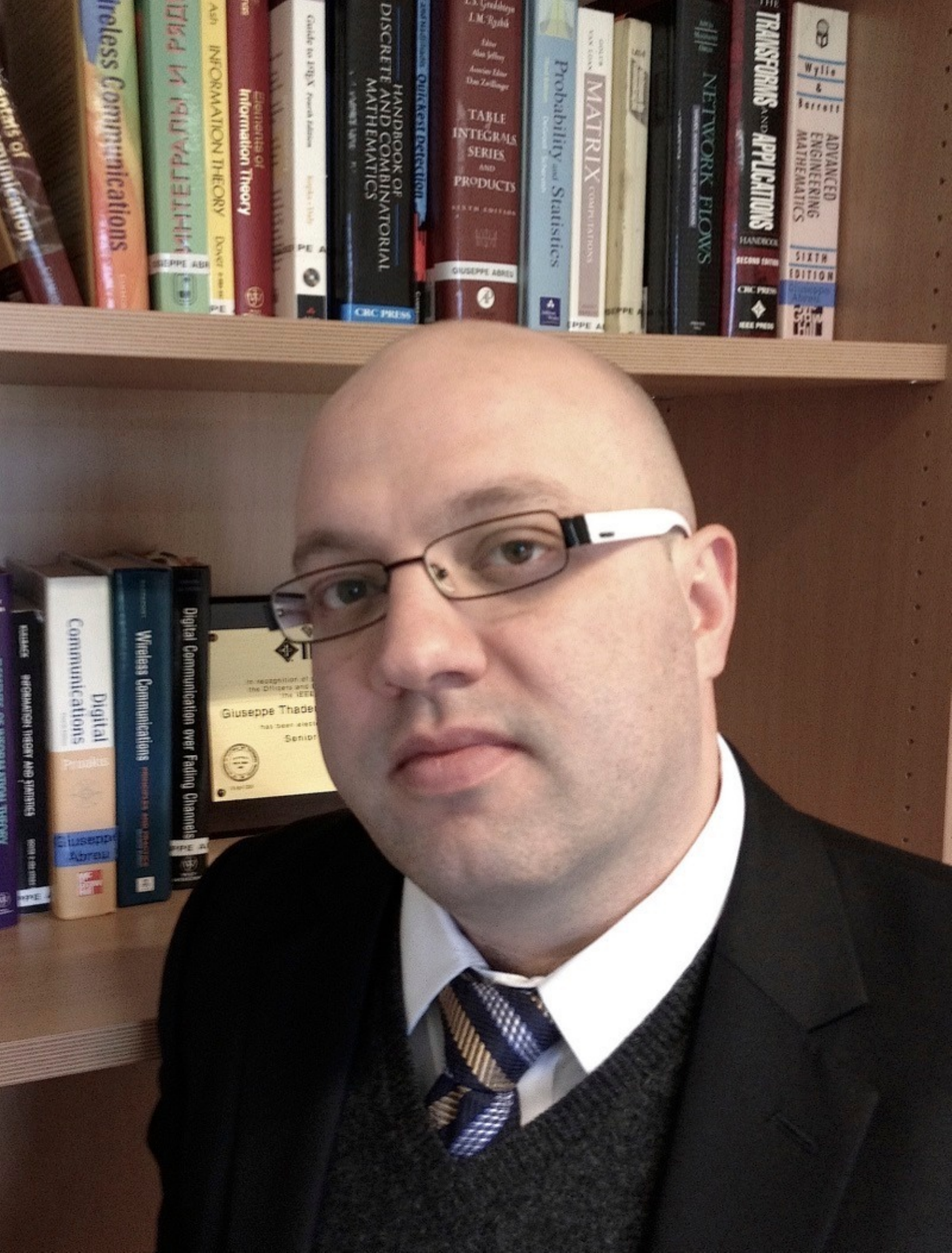}}]{Giuseppe Thadeu Freitas de Abreu} (Senior Member, IEEE) received the B.Eng. degree in electrical engineering and the specialization Latu Sensu
degree in telecommunications engineering from the
Universidade Federal da Bahia (UFBa), Salvador,
Bahia, Brazil in 1996 and 1997, respectively, and
the M.Eng. and D.Eng. degrees in physics, electrical,
and computer engineering from Yokohama National
University, Japan, in March 2001 and March 2004,
respectively. He was a postdoctoral fellow and later
an adjunct professor (docent) in statistical signal
processing and communications theory at the Department of Electrical and
Information Engineering, University of Oulu, Finland from 2004 to 2006 and
from 2006 to 2011, respectively. Since 2011, he has been a professor of
electrical engineering at Constructor University (formerly known as Jacobs
University), Bremen, Germany. From April 2015 to August 2018, he also
simultaneously held a full professorship at the Department of Computer
and Electrical Engineering, Ritsumeikan University, Japan. His research
interests include communications and signal processing, communications
theory, estimation theory, statistical modeling, wireless localization, cognitive radio, wireless security, MIMO systems, ultrawideband and millimeter
wave communications, full-duplex and cognitive radio, compressive sensing,
energy harvesting networks, random networks, connected vehicles networks,
electro-magnetic signal processing, quantum computing for signal processing,
metasurfaces for wireless systems and many other topics. Prof. Abreu has
received various awards and prestigious fellowships, including the Uenohara
Award from Tokyo University in 2000, the prestigious JSPS, Heiwa Nakajima,
and NICT (twice) fellowships in 2010, 2013, 2015, and 2018, as well
as Best Paper award at several international conferences, and of the Best
Paper award by the Japanese Chapter of the IEEE Signal Processing Society
in 2023. He served as an associate editor for the IEEE Transactions on
Wireless Communications from 2009 to 2014 and the IEEE Transactions
on Communications from 2014 to 2017; as an executive editor for IEEE
Transactions on Wireless Communications from 2017 to 2021, and as an
editor IEEE Communications Letters from 2021 to 2023. He is currently
serving as an editor to the IEEE Signal Processing Letters and to the IEEE
Open Journal of the Communications Society.
\end{IEEEbiography}
\begin{IEEEbiography}[{\includegraphics[width=1in,height=1.25in,clip,keepaspectratio]{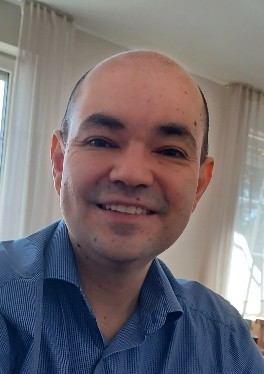}}]{David González G., PhD.}
(S'06-M'15-SM'18) has a master in Mobile Communications and PhD in Signal Theory and Communications from the Universitat Politècnica de Catalunya, Spain. He has served as post-doctoral fellow in Aalto University, Finland (2014-2017). He also served as Research Engineer with Panasonic Research and Development Center, Germany. Since 2018, David is with Continental Automotive Technologies, Germany, where he conducts and manages research projects focused on diverse aspects of vehicular communications (V2X), integrated sensing and communications (ISAC), and automotive applications for 5G-Advanced and 6G. David participates as delegate in 3GPP RAN1, 5GAA, and ETSI ISG ISAC. 
\end{IEEEbiography}
\begin{IEEEbiography}[{\includegraphics[width=1in,height=1.25in,clip,keepaspectratio]{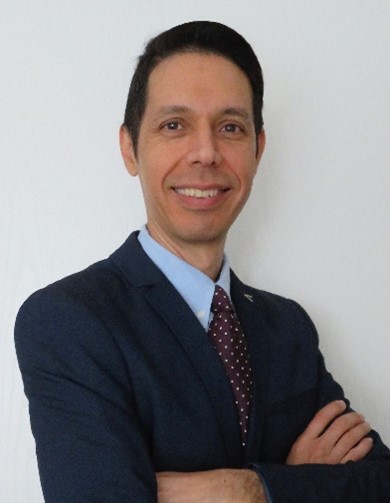}}]{Osvaldo Gonsa}
received the Ph.D. degree in electrical and computer engineering from Yokohama National University, Japan, in 1999, and the M.B.A. degree from the Kempten School of Business, Germany, in 2012. Since 1999 he has worked in research and standardization in the areas of core and radio access network. He is currently the Head of the Wireless Communications Technologies Group, Continental AG, Frankfurt, Germany. And since 2020 also serves as a member for the GSMA Advisory Board for automotive and the 6GKom Project of the German Federal Ministry of Education and Research.
\end{IEEEbiography}
}

\end{document}